\documentclass[twocolumn,twoside, trackchanges]{aastex701} 
\usepackage{appendix}
\usepackage{tabularx}
\usepackage{amsmath, amssymb}
\usepackage{rotating}
\usepackage{multirow}

\begin{document}

\title{Multiwavelength Study of Blue Straggler Stars in Tombaugh 2: Evidence for Binary Mass Transfer and Constraints on Cluster Dynamical State}

\correspondingauthor{D. Bisht}
\email{devendrabisht297@gmail.com}  

\author[orcid=0000-0002-8988-8434, sname=Bisht, gname=D.]{D. Bisht}\thanks{devendrabisht297@gmail.com}
\affiliation{Indian Centre For Space Physics
466, Barakhola, Singabari road, Netai Nagar, Kolkata, West Bengal, 700099}
\email{devendrabisht297@gmail.com}

\author[orcid=0000-0001-7359-3300, sname=Jiang, gname=D.]{Ing-Guey Jiang}\thanks{E-mail: jiang@phys.nthu.edu.tw}
\affiliation{Department of Physics and Institute of Astronomy, National Tsing-Hua University, Hsinchu 30013, Taiwan}
\email{jiang@phys.nthu.edu.tw}

\author[sname=Belwal, gname=K.]{K. Belwal}\thanks{E-mail: kuldeepbelwal1997@gmail.com}
\affiliation{Indian Centre For Space Physics
466, Barakhola, Singabari road, Netai Nagar, Kolkata, West Bengal, 700099}
\email{kuldeepbelwal1997@gmail.com}

\author[orcid=0000-0001-7940-3731, sname=Çınar, gname=Deniz Cennet]{D. C. Çınar}\thanks{E-mail: denizcennetcinar@gmail.com}
\affiliation{Programme of Astronomy and Space Sciences, Institute of Graduate Studies in Science, Istanbul University, Istanbul, 34116, Turkey}
\email{denizcdursun@gmail.com}

\author[orcid=0000-0002-4729-9316, sname=Dattatrey, gname=Arvind]{Arvind K. Dattatrey}
\affiliation{Indian Institute of Astrophysics, 560034 Bangalore, India}
\email{arvind@aries.res.in}

\author[orcid=0000-0002-6373-770X, sname=Rangwal, gname=Geeta]{Geeta Rangwal}
\affiliation{Aryabhatta Research Institute of Observational Sciences, Manora Peak, Nainital 263129, India}
\email{geetarangwal91@gmail.com}

\author[sname=Raj, gname=A.]{A. Raj}
\affiliation{Indian Centre For Space Physics
466, Barakhola, Singabari road, Netai Nagar, Kolkata, West Bengal, 700099}
\email{raj@example.com}

\author[orcid=0009-0003-8446-4557, sname=Biswas, gname=S.]{Shraddha Biswas}
\affiliation{Indian Centre For Space Physics
466, Barakhola, Singabari road, Netai Nagar, Kolkata, West Bengal, 700099}
\email{sbiswas@example.com}

\author[sname= Singh Bisht, gname=Mohit]{Mohit Singh Bisht}
\affiliation{Indian Centre For Space Physics
466, Barakhola, Singabari road, Netai Nagar, Kolkata, West Bengal, 700099}
\email{mohitsinghbisht742@gmail.com}

\author[sname=Durgapal, gname=Alok]{Alok Durgapal}
\affiliation{Center of Advanced Study, Department of Physics, D. S. B. Campus, Kumaun University, Nainital 263002, India}
\email{alokdurgapal@gmail.como}

\begin{abstract}

We present a focused multiwavelength study of blue straggler stars (BSSs) in the intermediate-age open cluster Tombaugh~2, located in the outer Galactic disk, to constrain the dominant formation pathways of BSSs in a low-density environment. Cluster members are identified using \textit{Gaia}~DR3 astrometry through a Gaussian Mixture Model, yielding a clean sample of high-probability members. Color--magnitude diagram analysis indicates an age of $\sim$1.74~Gyr. The radial surface density profile is well described by a King model, indicating a centrally concentrated overall structure, while the cluster exhibits only weak or no clear evidence of mass segregation among its stellar populations. We identify 26 BSS candidates and 2 YSS candidates. Spectral energy distributions constructed from ultraviolet, optical, and infrared photometry reveal that 9 BSSs ($\sim$32\%) exhibit significant ultraviolet excess, indicating an additional hot component. Binary SED decomposition identifies stripped companions with effective temperatures $T_{\rm eff} \approx (1.5$--$8)\times10^{4}\,\mathrm{K}$ and radii $R \approx 0.04$--$0.28\,R_\odot$, consistent with proto--white dwarfs, extremely low-mass pre--helium white dwarfs, and young hot remnants formed through recent mass transfer. A slight central concentration of BSSs, together with stripped companions, suggests that binary mass transfer is an important formation channel,  with no evidence for merger-driven formation. Multi-epoch VLT/FLAMES spectroscopy reveals radial-velocity variability in several systems, providing independent evidence for binarity.  Our results highlight that optical–infrared photometric analyses alone may fail to detect hot compact companions, while spectroscopy and ultraviolet observations provide complementary constraints, with ultraviolet data offering a direct probe of such companions in intermediate-age open clusters.

\end{abstract}

\keywords{\uat{Open star clusters}{1160} --- \uat{Blue straggler stars} {168} --- \uat{Binary Stars}{154}--- \uat{Spectral energy distribution}{2129}--- \uat{White Dwarf Stars}{1799}---\uat{Ultraviolet Astronomy}{1736}}


\section{Introduction} 

Star clusters provide well-defined stellar populations with common ages, distances, and chemical compositions, making them key laboratories for studies of stellar and Galactic evolution. Among their stellar constituents, BSSs are of particular interest, as they appear brighter and bluer than the main-sequence turn-off (MSTO), implying anomalously high masses for their host populations \citep{sandage1953color}. Their existence contradicts standard single-star evolution theory, which predicts that stars of a given age should simultaneously leave the main sequence. This discrepancy indicates that BSSs have undergone a mass-gain process that artificially extends their hydrogen-burning lifetimes \citep{ahumada2007, momany2007}.  Yellow straggler stars (YSSs) are objects located between the main-sequence turn-off (TO) and the red giant branch, often interpreted as evolved counterparts of blue stragglers \citep{leiner2016k2}.

The formation of BSSs is generally attributed to three main channels: (1) mass transfer in binary systems, (2) binary mergers (coalescence), and (3) stellar collisions, particularly in dense environments \citep{mccrea1964extended, hills1976stellar, leonard1989stellar}. In addition, triple-system evolution through Kozai–Lidov cycles can drive the inner binary toward mass transfer or merger, thereby contributing to these binary evolution channels  \citep{perets2009triple}. While direct collisions may occur in the dense cores of globular clusters, in the lower-density environments characteristic of open clusters (OCs), mass transfer in primordial binaries is considered the dominant mechanism \citep{mccrea1964extended, davies2004origins}. Even in low-density OCs, dynamical interactions can still alter binary systems. Binary–single and binary–binary encounters may lead to exchange events or the formation of dynamically assembled binaries \citep{Knigge2009, leigh2011analytic}. Depending on the evolutionary state of the donor star, mass transfer can occur during Case A (while the donor is still on the main sequence), Case B (after the exhaustion of core hydrogen), or Case C (after the onset of helium burning). These channels can produce a rejuvenated BSS together with a hot, compact remnant, such as a helium or carbon–oxygen white dwarf (WD), whose presence can be revealed through excess ultraviolet (UV) emission \citep{gosnell2014detection,subramaniam2018ultraviolet,li2019formation}.

Hot, compact companions emit strongly in the UV regime while contributing negligibly at optical wavelengths, making UV observations a powerful and direct diagnostic for identifying such systems and constraining the mass transfer history of recently formed BSSs \citep{sahu2019detection, Subramaniam2020}. While spectroscopic radial velocity (RV) monitoring can independently reveal binary companions through orbital motion, UV-based SED analysis provides a direct probe of their nature, particularly for hot, compact objects such as WDs that are otherwise difficult to detect at optical wavelengths. Distinguishing among different mass-transfer channels
(Case~A, Case~B, and Case~C) relies on identifying such companions, as indirect population-level indicators, including BSS luminosity relative to the TO, color--magnitude diagram (CMD) position, and radial distribution within the cluster (e.g., \citealt{lanzoni2007, rain2021new}) cannot uniquely constrain the mass transfer history of individual systems. We note that Case~A mass transfer may also result in a blue straggler with a main-sequence companion, which does not exhibit UV excess and may therefore remain undetected in UV-based analyses. Multiwavelength SED analysis is widely used to characterize these systems. In recent years, this approach has been extensively and successfully applied to identify WD companions across a variety of OCs, offering direct observational constraints on BSS formation scenarios \citep[e.g.,][]{gosnell2015implications, jadhav2019blue, panthi2022analysis, Jadhav2024, Saketh2024, Pal2024}.

Tombaugh 2 is an intermediate-age open cluster located in the outer Galactic disk. Since stellar encounter rates depend primarily on cluster density rather than Galactocentric distance, the relatively low density of Tombaugh 2 reduces the frequency of stellar interactions. This makes it a suitable environment for examining binary evolution as the dominant formation channel of BSSs, while also allowing us to assess whether the BSS population shows any imprint of dynamical processes, such as mass segregation or variations in radial distribution. It hosts a significant population of BSSs \citep{andreuzzi2011old, rain2021new, rao2023determination}. Its age ($\sim$1.74~Gyr) places it in a regime where long-period binary evolution can proceed efficiently. Previous photometric and spectroscopic studies have established its age, metallicity, and kinematics and have suggested a possible connection to outer-disk substructures, such as the Galactic Anticenter Stellar Structure \citet{frinchaboy2008tombaugh, villanova2010metallicity}. Crucially, Tombaugh~2 benefits from available UV observations \citep{Siegel2019}, enabling a direct search for hot companions that is not possible for many outer-disk OCs.

In this work, we investigate binary mass transfer as the primary formation mechanism of BSSs in the low-density open cluster Tombaugh~2. We identify hot, compact companions through UV excess emission to constrain the nature of the mass-transfer process and distinguish among different mass-transfer channels. In addition, we examine the radial distribution of BSSs and estimate dynamical friction timescales to assess whether the observed population shows signatures of dynamical effects on the BSS population, such as mass segregation. We combine Gaia DR3 astrometry with UV, optical, and infrared photometry to construct a clean sample of cluster members. We perform multiwavelength SED analysis to identify UV excess and use multi-epoch spectroscopic data to investigate radial-velocity variability. By linking these observational signatures with the cluster’s structural and dynamical properties, we find evidence for a significant binary fraction among BSSs in Tombaugh 2 and direct evidence for recent or ongoing mass-transfer activity in a subset of systems. These results provide empirical support for binary evolution as a major formation channel of BSSs in outer-disk OCs.

The paper is organised as follows. Section~\ref{sec:data} describes the data sets used in this study. Section~\ref{sec: membership} presents the membership determination. Section~\ref{sec:CMD} examines the color–magnitude diagram and spatial distribution of BSSs. Section~\ref{sec: sed} presents the SED analysis and identification of candidate binary systems. Section~\ref{sec:RV_analysis} discusses the RV analysis and its implications for membership and binarity. Section~\ref{sec: rdp} discusses the structural parameters derived from radial surface density profiles. Section~\ref{sec: bss_dynamics} explores the orbital properties of the cluster, and Section~\ref{sec: conclusion_summary} summarises our conclusions.

\section{ Data}
\label{sec:data}

In this study, we utilise multiwavelength photometric data spanning the UV, optical, and infrared (IR) regimes to characterize the properties of BSS and YSS candidates in Tombaugh 2. All data used in this work are drawn from archival space and ground-based surveys.

\subsection{Ultraviolet (UV) Data}
UV flux measurements, critical for identifying BSS candidates and potential hot companions, were obtained from the UV and Optical Telescope (Swift/UVOT) aboard the Neil Gehrels Swift Observatory. Hot stars emit a significant fraction of their flux in the UV. UVOT is a modified 30 cm Ritchey-Chr\'{e}tien telescope offering a wide field of view of $17^{\prime}\times17^{\prime}$ \citep{roming2000ultraviolet, gehrels2004swift}. For the analysis presented herein, UV photometry was primarily sourced from the $UVM2$ (effective wavelength $\lambda_{eff} \approx 2246$ \AA) and $UVW2$ ($\lambda_{eff} \approx 2086$ \AA) filters. 

Rather than processing raw images, we used calibrated flux measurements from the Level~3 point-source catalogs of \citet{Siegel2019}, archived in the \textit{Swift}/UVOT database. For Tombaugh~2, the available UV data provide a total integrated exposure time of 1979~s per filter, sufficient for detecting hot components in BSS systems. UV observations from facilities such as HST, UVIT, and \textit{Swift}/UVOT have been widely used to identify hot white-dwarf companions in blue straggler systems \citep{gosnell2015implications, sindhu2019uvit, jadhav2021high, rao2022characterization}.

\subsection{Optical Photometry and Spectroscopic Data}

The core optical photometry for our study was compiled from several surveys. Astrometric data (positions $\alpha$, $\delta$; trigonometric parallaxes $\varpi$; and proper motions $\mu_{\alpha} \cos\delta$, $\mu_{\delta}$) necessary for cluster membership determination, along with fundamental optical photometry, were sourced from the Gaia mission's third Data Release \citep{GaiaCollaboration2021,collaboration2023gaia}. Gaia DR3 provides high-precision measurements for about 1.46 billion sources, including photometry in the $G$ ($\lambda_{eff} \approx 673$ nm), $G_{BP}$ ($\lambda_{eff} \approx 532$ nm), and $G_{RP}$ ($\lambda_{eff} \approx 797$ nm) bands \citep{brown2021gaia}. To ensure high-quality astrometric solutions, we selected only those stars with a Renormalised Unit Weight Error (RUWE) $\leq$~1.4 \citep{lindegren2021gaia}. We limited the sample to stars with $G$-band magnitudes $\leq 20.1$ to ensure reliable precision. Broadband optical data were also obtained from the Panoramic Survey Telescope and Rapid Response System (Pan-STARRS1) Data Release 2, which incorporates measurements in the \texttt{g}, \texttt{r}, \texttt{i}, \texttt{z}, and \texttt{y} filters \citep{hodapp2004design, stubbs2010precise}.
To further extend the photometric baseline, data from the SkyMapper Southern Survey (SMSS) Data Release 4 were included, utilizing the \texttt{u}, \texttt{v}, \texttt{g}, \texttt{r}, \texttt{i}, and \texttt{z} filters \citep{onken2024skymapper}.

The spectroscopic data analyzed in this work were retrieved from the ESO Science Archive Facility as part of the Phase 3 data release\footnote{\url{https://archive.eso.org/scienceportal/}}. The observations were carried out using the FLAMES (Fiber Large Array Multi-Element Spectrograph) instrument \citep{Pasquini2002}, mounted on the UT2 (Kueyen) telescope of the VLT at the ESO Paranal Observatory. The data were acquired over five distinct observing epochs between 2006 February 20 and March 25 (see Table \ref{tab:obs_log}), under the program ID 076.D-0220 (PI: S. Randich).

\begin{table}[h]
\centering
\caption{Log of spectroscopic observations.}
\label{tab:obs_log}
\begin{tabular}{c c c}
\hline \hline
Epoch (ID) & Date (UT) & HJD (days) \\
 & & (-2,450,000) \\
\hline
1 & 2006 Feb 20 04:00 & 3786.67 \\
2 & 2006 Mar 06 02:30 & 3800.60 \\
3 & 2006 Mar 20 02:30 & 3814.60 \\
4 & 2006 Mar 24 02:30 & 3818.60 \\
5 & 2006 Mar 25 00:25 & 3819.52 \\
\hline
\end{tabular}
\end{table}

The observations used the GIRAFFE spectrograph in the MEDUSA mode with the HR15N high-resolution grating. This setup covers the wavelength range of $\lambda\lambda$ 6440--6820 \AA\ (central wavelength 6650 \AA) with a resolving power of $R \sim 17,000$. This spectral window is particularly well suited to our analysis, as it encompasses the H$\alpha$ line ($\lambda$ 6562.8 \AA), a robust diagnostic for RV measurements. A uniform integration time of 2775 s was adopted for all exposures to ensure consistent data quality.

We used the reduced one-dimensional spectra from the ESO Phase 3 pipeline. The standard reduction procedures applied include bias subtraction, flat-field correction, fiber extraction, wavelength calibration, and sky subtraction using dedicated sky fibers. We performed a quality-control check of the individual spectra by measuring the signal-to-noise ratio ($S$/$N$) at the continuum level. The dataset exhibits S/N values ranging from 16.1 to 40.1, ensuring sufficient spectral quality for precise RV determinations. These $S$/$N$ values are sufficient to achieve RV precision of a few km $s^{-1}$, adequate for detecting the variability expected from interacting binaries.  The typical uncertainties in the measured RVs are of the order of $\sim$1--3 km s$^{-1}$, depending on the signal-to-noise ratio of the spectra.

\subsection{Infrared (IR) Data}
Near-infrared (NIR) and mid-infrared (MIR) data were used to constrain the cooler regime of the stars' spectral energy distribution (SED). NIR photometry was sourced from the Two Micron All Sky Survey (2MASS) catalog, using data from the \texttt{J} (1.24 $\mu$m), \texttt{H} (1.66 $\mu$m), and \texttt{$K_s$} (2.16 $\mu$m) bands \citep{skrutskie2006two}. MIR data were obtained from the AllWISE data release of the Wide-field Infrared Survey Explorer (WISE) mission, prioritizing the \texttt{W1} (3.35 $\mu$m) and \texttt{W2} (4.60 $\mu$m) bands \citep{wright2010wide}, which generally offer higher signal-to-noise ratios.

\begin{figure*}[t]
    \centering
    \includegraphics[width=1\linewidth]{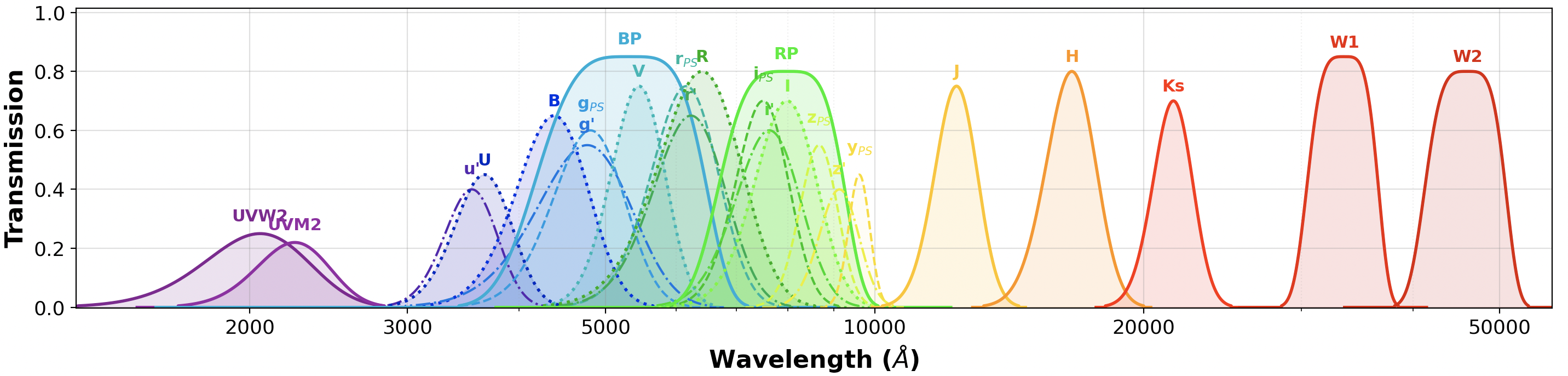}
    \caption{Normalized filter transmission curves illustrating the wavelength coverage of the UV, optical, and infrared data used in the SED analysis.}
\label{fig:transmission_filters}
\end{figure*}

All multi-wavelength photometric data points used to construct the SEDs of the target stars were compiled using the Virtual Observatory SED Analyzer (VOSA) tool, with filter transmission curves adopted from the Spanish Virtual Observatory Filter Profile Service\footnote{\url{https://svo2.cab.inta-csic.es/theory/fps/}} (SVO-FPS) to illustrate the wavelength coverage in Figure~\ref{fig:transmission_filters}.

\section{Astrometric Membership Determination}
\label{sec: membership}

To reliably identify cluster members in Tombaugh~2, we employed the method of \citet{agarwal2021ml} in the astrometric parameter space. GMM is a probabilistic clustering technique that models the data as a collection of Gaussian distributions and assigns membership probabilities based on the likelihood of belonging to each cluster \citep{mclachlan2000finite, reynolds2009gaussian}. GMM effectively separates cluster and field populations in astrometric space, although its performance depends on the degree of overlap between the two distributions.

The cluster is located at a low Galactic latitude ($b$ = $-6.9^{\circ}$), where field-star contamination is significant.  As a result, the adopted method is particularly effective for intermediate-age, sparsely populated clusters such as Tombaugh 2, where classical selection techniques are prone to contamination.  Robust cluster membership is essential for interpreting the blue straggler population, particularly when identifying UV excess sources that may otherwise be contaminated by field objects. The GMM-based astrometric selection adopted here ensures a clean and reliable member sample \citep{gao2018machine, qiu2024deeper, belwal2026time}. Cluster membership is determined solely using Gaia DR3 astrometric parameters (proper motion and parallax), without incorporating photometric selection criteria.

Afterward, we refined the input dataset following the procedure outlined in our previous work by \citet{belwal2025unveiling}. Briefly, we retrieved \textit{Gaia}~DR3 sources within a circular region centered on the cluster, retaining only stars with complete five-parameter astrometric solutions and valid photometry in the $G$, $G_{BP}$, and $G_{RP}$ bands. Quality cuts were applied by selecting stars with positive parallaxes, proper motion uncertainties $\leq \epsilon_{\mu}$, and RUWE $\leq 1.4$ \citep{lindegren2021gaia}. We note that strict RUWE constraints can exclude unresolved binary systems; therefore, this criterion was applied cautiously to balance astrometric quality and sample completeness. We adopted a proper-motion uncertainty threshold of $\epsilon_{\mu} \leq 0.5$ mas\,yr$^{-1}$ to ensure reliable astrometric measurements while retaining a statistically significant sample, consistent with typical Gaia-based open cluster studies. To minimize field contamination, an initial kinematic pre-selection was made in the proper motion–parallax plane using the $k$-Nearest Neighbours (kNN) algorithm \citep{cover1967nearest} to estimate the cluster centroid in the three-dimensional space ($\mu_{\alpha}^{*}$, $\mu_{\delta}$, $\varpi$).  We verified that sources with near-zero parallaxes correspond to background field stars and are not included in the final membership sample, as ensured by the combined kNN and GMM selection criteria. The proper motion and parallax ranges of the kNN-selected sources are [0 to 0.4] mas for ($\varpi$), [-1.3 to 0.3] mas yr$^{-1}$ for $\mu_{\alpha}^{*}$, and [0.70 to 2.1] mas yr$^{-1}$ for $\mu_{\delta}$. The field-of-view radius for the sample data is 10 arcmin. The refined dataset was then normalized and analyzed using a two-component GMM implemented through the Expectation–Maximization algorithm \citep{dempster1977maximum, mclachlan2000finite}. Each star was assigned a soft membership probability, and only stars with $p \geq p_\mathrm{min}$ were considered bona fide cluster members for subsequent analysis. The mean astrometric parameters derived from these members agree well with recent literature values \citep{Cantat-Gaudin20} and \citep{hunt2023improving}.

In total, 562 stars in Tombaugh~2 were identified as probable members with membership probabilities greater than 70$\%$. Of these, 454 stars were successfully cross-matched with the \citet{hunt2023improving} catalog, confirming the consistency of our member selection. The mean proper motion values derived from our GMM analysis are $\mu_{\alpha} = -0.51 \pm 0.11$~mas~yr$^{-1}$ and $\mu_{\delta} = 1.41 \pm 0.12$~mas~yr$^{-1}$. The mean motion obtained for Tombaugh~2 aligns closely with the values of \citet{hunt2023improving} ($\mu_{\alpha} = -0.51$~mas~yr$^{-1}$, $\mu_{\delta} = 1.41$~mas~yr$^{-1}$) and shows only a slight offset compared to \citet{Cantat-Gaudin20} ($\mu_{\alpha} = -0.46$~mas~yr$^{-1}$, $\mu_{\delta} = 1.35$~mas~yr$^{-1}$). The consistency of our mean astrometric parameters with these independent studies strongly supports the robustness of the adopted GMM-based membership determination.

\begin{figure*}
    \includegraphics[width=18cm,height=6.5cm]{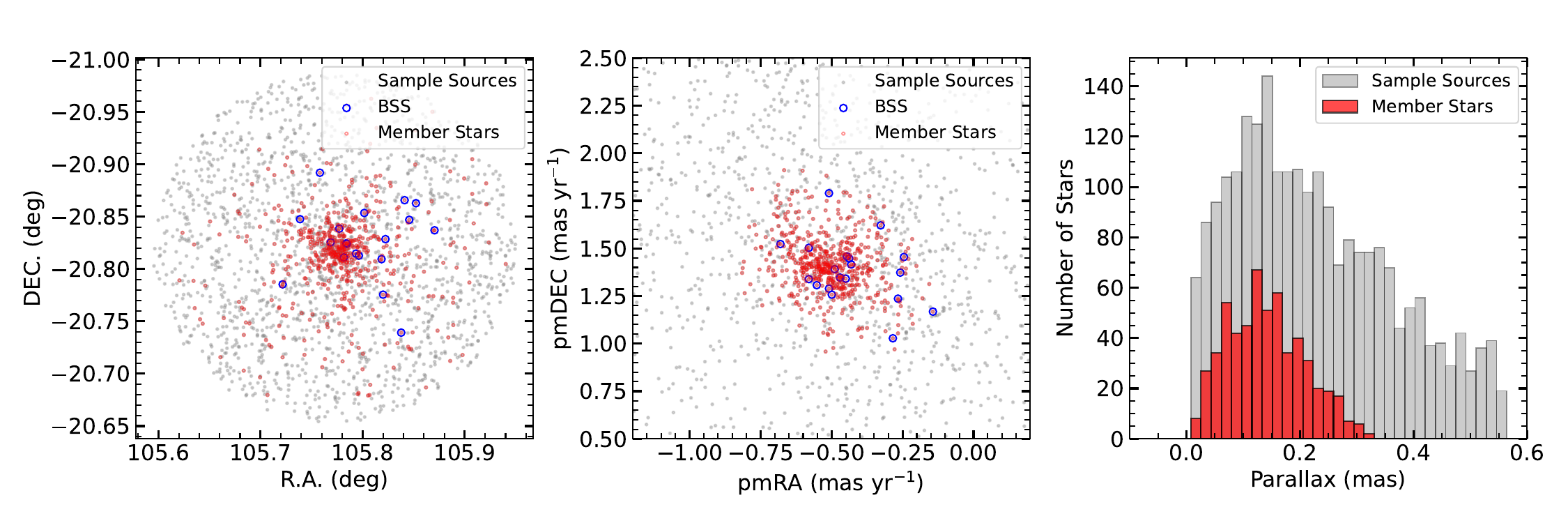}
    \caption{Spatial, kinematic, and parallax distributions for Tombaugh~2. Red points represent member stars identified via the GMM analysis, blue circles mark BSSs, and grey points correspond to all stars. The proper motion and parallax panels indicate distinct clustering of members, validating the adopted membership determination. Cluster membership is determined using a combination of kNN pre-selection and GMM classification.}
    \label{fig: members_both}
\end{figure*}

After estimating the membership probabilities using our kNN+GMM selection method, we examined the spatial, kinematic, and parallax distributions of the identified members. Figure~\ref{fig: members_both} presents these diagnostic diagrams for Tombaugh~2. The spatial, kinematic, and parallax distributions show a clear separation between cluster members and field stars, confirming the robustness of the adopted membership selection.

The distance to the cluster was estimated from the mean parallax of the probable members. To address the known nonlinearity and asymmetry in parallax uncertainties, we used the probabilistic approach proposed by \citet{bailer2018estimating}. This method provides distance estimates using a Galactic prior that accounts for the spatial distribution of stars. It offers a more realistic treatment of parallax uncertainties than the simple inverse-parallax conversion, especially for distant or faint stars. For the present sample, the mean parallax is $0.10 \pm 0.08$~mas, corresponding to a distance estimate of $7.09^{+0.50}_{-0.44}$~kpc after applying the Bailer-Jones criteria. This value is consistent, within uncertainties, with the distances reported by \citet{Cantat-Gaudin20} and \citet{hunt2023improving}. The close agreement between our derived distance and that from recent astrometric studies reinforces the reliability of the adopted membership selection and parallax correction procedure. This consistency also indicates that systematic errors in parallax zero-point or astrometric calibration are minimal for this distant open cluster.

\section{Color--Magnitude Diagram and Blue Straggler Population}\label{sec:CMD}

\begin{figure}
\centering 
\includegraphics[width=\linewidth]{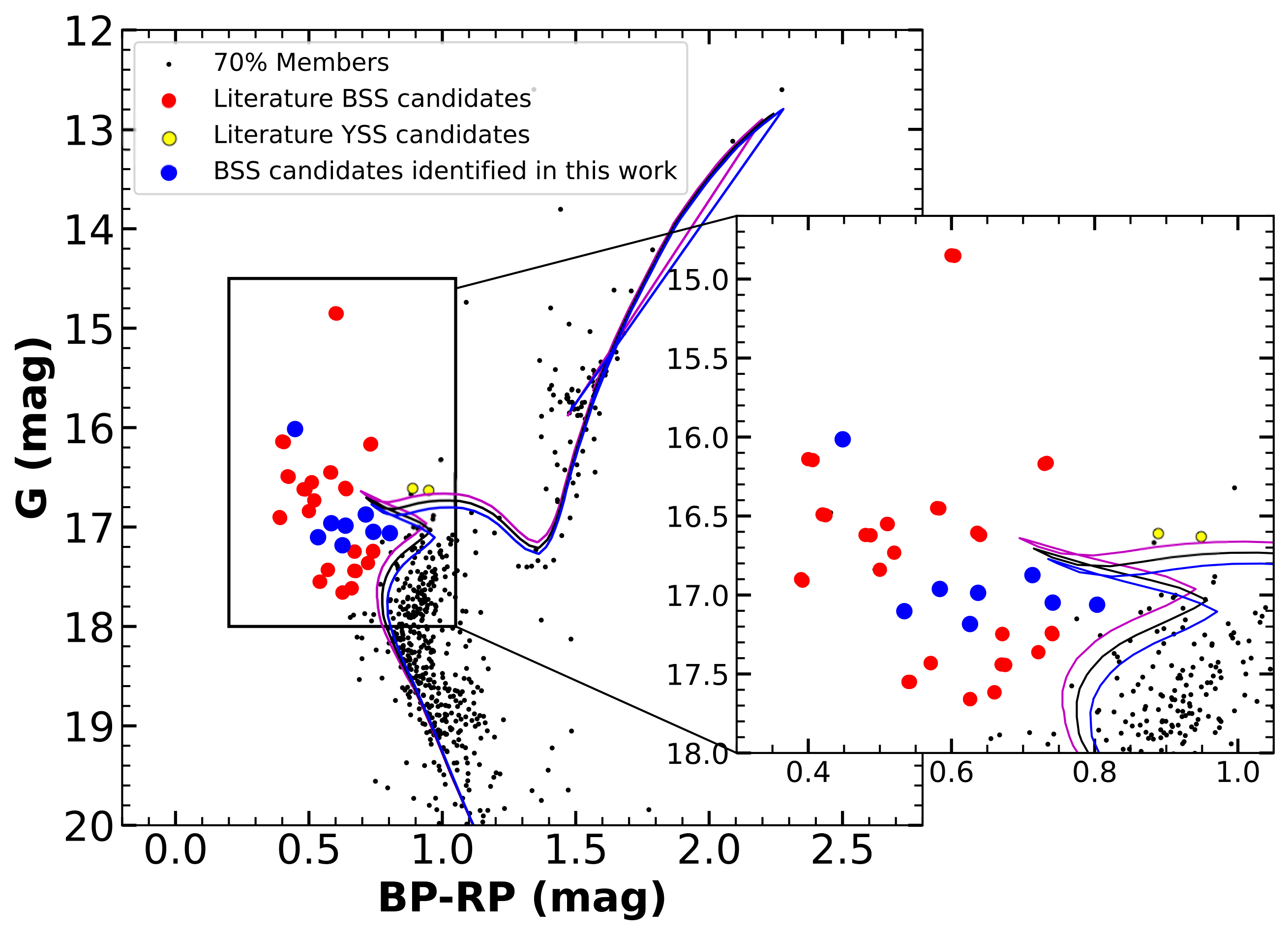}
\caption{CMD of the cluster Tombaugh~2, plotted in the $G$ versus $(\mathrm{BP}-\mathrm{RP})$ plane. Black points represent cluster members with membership probabilities $\geq 70\%$, selected based on astrometric criteria. The solid blue curves show the best-fitting PARSEC isochrones corresponding to the derived cluster parameters. The fitted isochrones correspond to $\log(\mathrm{age/yr}) = $9.22, 9.24, and 9.26, with the central value representing the best fit and the adjacent curves illustrating the age uncertainty. Red solid dots represent the BSS candidates reported in literature, while the filled blue dots mark the BSS candidates identified in this study.}
\label{fig:cmd}
\end{figure}

Figure~\ref{fig:cmd} presents the Gaia-based color--magnitude diagram (CMD) of Tombaugh~2, constructed using stars with membership probabilities $\geq 70\%$  as determined from the astrometric membership analysis. The CMD exhibits a well-defined cluster sequence extending from the main sequence (MS) through the TO region to the red giant branch (RGB). The MSTO is located at $G \approx 17.5$~mag, and the CMD displays a prominent red clump, consistent with an age of $\log(\mathrm{age/yr}) = 9.24 \pm 0.02$ ($\sim$1.74~Gyr). PARSEC isochrones\footnote{\url{http://stev.oapd.inaf.it/cmd}} \citep{Bressan2012} with metallicities $Z$ = 0.007 $\pm$ 0.001 appropriate for outer-disk OCs provide a good match to the observed sequences, supporting a moderately metal-poor population. These parameters are consistent with previous photometric and spectroscopic studies of Tombaugh~2 \citep{hunt2023improving, Cantat-Gaudin20}. A distinct population of stars is observed above and blueward of the MS TO, which we identify as BSS candidates. We also identify a subset of stars slightly offset from the canonical BSS region in the color–magnitude diagram; following previous studies, we include these as BSS candidates. In general, all stars located in the BSS region of the CMD are treated as BSS candidates in this study.

We determined the distance modulus and the color excess $E(BP-RP)$ to be $15.50 \pm 0.10$ mag and $0.45$ mag, respectively, from the best-fitting isochrone solution. To assess the reliability of the fits, we applied the reduced chi-square test for isochrone fitting following the methodology of \citet{valle2021goodness}. The optimal values of the distance modulus and color excess were obtained by minimizing the $\chi^{2}$ statistic, which quantifies the difference between the observed and model CMDs while accounting for observational uncertainties. The resulting reduced chi-square value is 0.75, indicating a good fit. The isochrone fitting is optimized around the main-sequence and TO regions, and minor deviations in the red giant branch and red clump are within expected uncertainties. We used the derived value of $E(BP-RP)$ to estimate $E(B-V)$ using the transformation relation $E(B-V) = 0.72 \times E(BP-RP)$ \citep{casagrande2018use}, obtaining $E(B-V) = 0.33 \pm 0.08$ mag. We adopted a standard total-to-selective extinction ratio of $R_{V} = 3.1$ \citep{cardelli1989relationship}. Using the value of $R_{V}$, we derived $A_{V} = 1.04 \pm 0.10$ mag. These values are consistent with those reported by \citet{rao2023determination}. We adopted the cluster's metallicity from high-resolution spectroscopic data from the APOGEE survey \citep{majewski2017apache}. Specifically, nine stars located within the cluster region with membership probabilities greater than 70$\%$ were selected. A Gaussian fit to the metallicity distribution of these member stars yields a mean metallicity of [Fe/H] = $-0.35$ $\pm$ 0.04 dex. This result agrees well within uncertainties with previous determinations of [Fe/H] = $-0.31$ dex reported by \citet{villanova2010metallicity}, confirming the reliability of our estimate.  

We cross-matched our BSS candidates with the catalogs of \citet{jadhav2021high} and \citet{rain2021new}, identifying 17 and 18 common stars, respectively, of which 16 stars are common to both catalogs. Literature BSS candidates are overplotted in Figure~\ref{fig:cmd} (red solid dots), along with the newly identified BSS candidates from this study (blue filled circles). The comparison shows a partial overlap between our identified BSS candidates and those reported in previous studies, indicating general consistency while also highlighting additional candidates identified in this work. This difference likely arises from variations in selection criteria, photometric depth, and membership determination methods across studies. The CMD enables a consistent selection of BSS candidates for subsequent analysis.

\begin{figure}
    \centering
    \includegraphics[width=\linewidth]{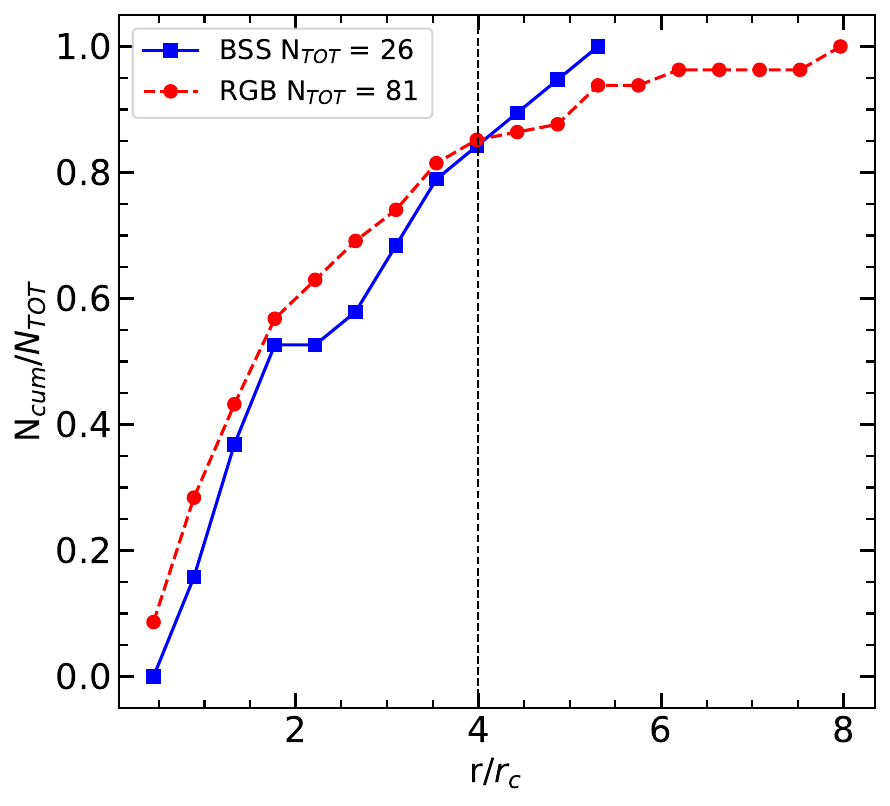}
    \caption{Cumulative radial distributions of BSSs (blue squares) and RGBs (red circles) in the cluster. The normalized cumulative fraction, $N_{\rm cum}/N_{\rm tot}$, is shown as a function of projected radial distance expressed in units of the cluster core radius ($r_c$). The vertical dashed line indicates the radial extent within which the BSS population appears more centrally concentrated than the RGB stars; beyond this radius, the cumulative distributions of the two populations converge.}
    \label{fig: radial_distribution}
\end{figure}

To construct the cumulative radial distributions, we selected the RGB stars as the reference population, following the approach of \citet{vaidya2020blue}. RGB stars are adopted as the reference population due to their well-defined evolutionary status and lower photometric incompleteness compared to main-sequence stars. We then derived the cumulative radial distributions of the BSS and RGB stars for this cluster. Figure~\ref{fig: radial_distribution} shows the resulting distributions, where the normalized cumulative number of stars is plotted along the y-axis and the radial distance, expressed in units of the core radius ($r_c$), is shown on the x-axis. To ensure a meaningful comparison between the BSS and RGB populations, only stars within the same magnitude range were considered, as indicated in Figure~\ref{fig: radial_distribution}. While the two distributions appear similar at $\approx 4.0\,r_c$ of the cluster, the BSS population shows only a slight central concentration within the inner regions; however, beyond $\approx 4.0\,r_c$, its distribution becomes comparable to that of RGB stars, indicating no clear evidence of strong dynamical segregation. Furthermore, a Kolmogorov–Smirnov test rejects the null hypothesis that the BSS and RGB populations are drawn from the same parent distribution at a confidence level of $\geq 95\%$. However, this statistical difference does not necessarily imply full mass segregation; rather, it reflects differences in their spatial distributions. Overall, this indicates weak or absent mass segregation and is consistent with an incompletely dynamically relaxed cluster, in agreement with its structural properties (see Section~\ref{sec: rdp}) and with the understanding that the relation between BSS segregation and dynamical age is not always straightforward \citep{vaidya2020blue, balan2024dynamical}.

\section{Spectral Energy Distribution Analysis of Blue Straggler Stars}
\label{sec: sed}

We analyzed the SED of stars identified as BSS candidates. \textit{Swift}/UVOT detections are available for 19 BSS candidates, while none of the YSSs are detected in the UV bands. Among the 26 identified BSS candidates, only these 19 sources have UV detections and are therefore included in the SED analysis. One BSS (BSS~19) was not included in the SED analysis due to the lack of sufficient photometric coverage. The SED fitting was performed using the Virtual Observatory SED Analyzer\footnote{\url{https://svo2.cab.inta-csic.es/theory/vosa}} (VOSA; \cite{Bayo2008}), an online tool designed for automated retrieval and fitting of multi-band photometric data. The input photometry, compiled as described in Section~\ref{sec:data}, spans from the UV (Swift/UVOT) through the optical (Gaia~DR3, Pan-STARRS, SkyMapper, and TESS) to the infrared (2MASS, WISE). UV data listed in Table~\ref{tab:bss_yss_fluxes} were used solely to identify the presence of secondary components, while the multi-band SED fitting was performed using only optical and IR data points.

For each source, we provide the equatorial coordinates ($\alpha$, $\delta$), the cluster distance, and the corresponding $V$-band extinction. We calculate a visual extinction of $A_{\rm V} = 1.04$~mag and a cluster distance of $d = 7.09^{+0.50}_{-0.44}$~kpc, determined from the distance approach provided by \citet{bailer2018estimating}. VOSA automatically queried multiple photometric archives within a $5\arcsec$ radius of the input position to extract flux measurements. Each field was visually inspected using Aladin\footnote{\url{https://aladin.u-strasbg.fr/}} to identify possible contamination from nearby stars within the retrieval radius; no significant blending was found. Interstellar extinction corrections were applied by VOSA using the extinction prescriptions of \citep{Fitzpatrick1999, Indebetouw2005}, assuming a standard total-to-selective ratio of $R_V = 3.1$.

The extinction-corrected SEDs were initially fitted with single-star atmosphere models from the Kurucz ODFNEW/NOVER grid \citep{Castelli1997, Castelli2003}. The parameter space covered $T_{\rm eff} = 5000$–$15000$~K. The best-fitting model was determined by minimizing the reduced chi-square statistic ($\chi^2_\nu$) between the observed and synthetic fluxes, with the synthetic fluxes computed by convolving the theoretical spectra with the adopted filters' response functions. In addition to $\chi^2_\nu$, we used the visual goodness-of-fit parameter (bolometric) ($V_{\rm gfb}$) provided by VOSA \citep{Bayo2008}, which defines a modified reduced $\chi^2$ by enforcing $\sigma(F_{\rm obs}) > 0.1\times F_{\rm obs}$ to mitigate the effect of underestimated photometric errors. According to the literature \citep[e.g.,][]{Jimenez-Esteban2021, Rebassa-Mansergas2019, Solano2021, Zeng2025}, models with $V_{\rm gfb} <$  10–15 are generally considered to provide good fits. Following this convention, only sources meeting this criterion were retained for further analysis.

\figsetstart
\figsetnum{1}
\figsettitle{SED fitting for the single-component models of the BSSs, YSSs, and BSS candidates. The complete set of 19 figures is provided as online-only supplementary material.}

\figsetgrpstart
\figsetgrpnum{1.1}
\figsetgrptitle{SED for BSS 12}
\figsetplot{1_T2.png}
\figsetgrpnote{SED for BSS 12}
\figsetgrpend

\figsetgrpstart
\figsetgrpnum{1.2}
\figsetgrptitle{SED for BSS 9}
\figsetplot{2_T2.png}
\figsetgrpnote{SED for BSS 9}
\figsetgrpend

\figsetgrpstart
\figsetgrpnum{1.3}
\figsetgrptitle{SED for BSS 6}
\figsetplot{3_T2.png}
\figsetgrpnote{SED for BSS 6}
\figsetgrpend

\figsetgrpstart
\figsetgrpnum{1.4}
\figsetgrptitle{SED for BSS 11}
\figsetplot{4_T2.png}
\figsetgrpnote{SED for BSS 11}
\figsetgrpend

\figsetgrpstart
\figsetgrpnum{1.5}
\figsetgrptitle{SED for BSS 8}
\figsetplot{5_T2.png}
\figsetgrpnote{SED for BSS 8}
\figsetgrpend

\figsetgrpstart
\figsetgrpnum{1.6}
\figsetgrptitle{SED for BSS 25}
\figsetplot{6_T2.png}
\figsetgrpnote{SED for BSS 25}
\figsetgrpend

\figsetgrpstart
\figsetgrpnum{1.7}
\figsetgrptitle{SED for YSS 1}
\figsetplot{7_T2.png}
\figsetgrpnote{SED for YSS 1}
\figsetgrpend

\figsetgrpstart
\figsetgrpnum{1.8}
\figsetgrptitle{SED for BSS 21}
\figsetplot{8_T2.png}
\figsetgrpnote{SED for BSS 21}
\figsetgrpend

\figsetgrpstart
\figsetgrpnum{1.9}
\figsetgrptitle{SED for BSS 24}
\figsetplot{9_T2.png}
\figsetgrpnote{SED for BSS 24}
\figsetgrpend

\figsetgrpstart
\figsetgrpnum{1.10}
\figsetgrptitle{SED for BSS 20}
\figsetplot{10_T2.png}
\figsetgrpnote{SED for BSS 20}
\figsetgrpend

\figsetgrpstart
\figsetgrpnum{1.11}
\figsetgrptitle{SED for BSS 1}
\figsetplot{11_T2.png}
\figsetgrpnote{SED for BSS 1}
\figsetgrpend

\figsetgrpstart
\figsetgrpnum{1.12}
\figsetgrptitle{SED for BSS 18}
\figsetplot{12_T2.png}
\figsetgrpnote{SED for BSS 18}
\figsetgrpend

\figsetgrpstart
\figsetgrpnum{1.13}
\figsetgrptitle{SED for BSS 19}
\figsetplot{13_T2.png}
\figsetgrpnote{SED for BSS 19}
\figsetgrpend

\figsetgrpstart
\figsetgrpnum{1.14}
\figsetgrptitle{SED for BSS 14}
\figsetplot{14_T2.png}
\figsetgrpnote{SED for BSS 14}
\figsetgrpend

\figsetgrpstart
\figsetgrpnum{1.15}
\figsetgrptitle{SED for BSS 17}
\figsetplot{15_T2.png}
\figsetgrpnote{SED for BSS 17}
\figsetgrpend

\figsetgrpstart
\figsetgrpnum{1.16}
\figsetgrptitle{SED for YSS 2}
\figsetplot{16_T2.png}
\figsetgrpnote{SED for YSS 2}
\figsetgrpend

\figsetgrpstart
\figsetgrpnum{1.17}
\figsetgrptitle{SED for BSS 4}
\figsetplot{17_T2.png}
\figsetgrpnote{SED for BSS 4}
\figsetgrpend

\figsetgrpstart
\figsetgrpnum{1.18}
\figsetgrptitle{SED for BSS 2}
\figsetplot{18_T2.png}
\figsetgrpnote{SED for BSS 2}
\figsetgrpend

\figsetgrpstart
\figsetgrpnum{1.19}
\figsetgrptitle{SED for BSS 16}
\figsetplot{19_T2.png}
\figsetgrpnote{SED for BSS 16}
\figsetgrpend

\figsetend

\begin{figure}[ht!]
    \centering
    \includegraphics[width=1\linewidth]{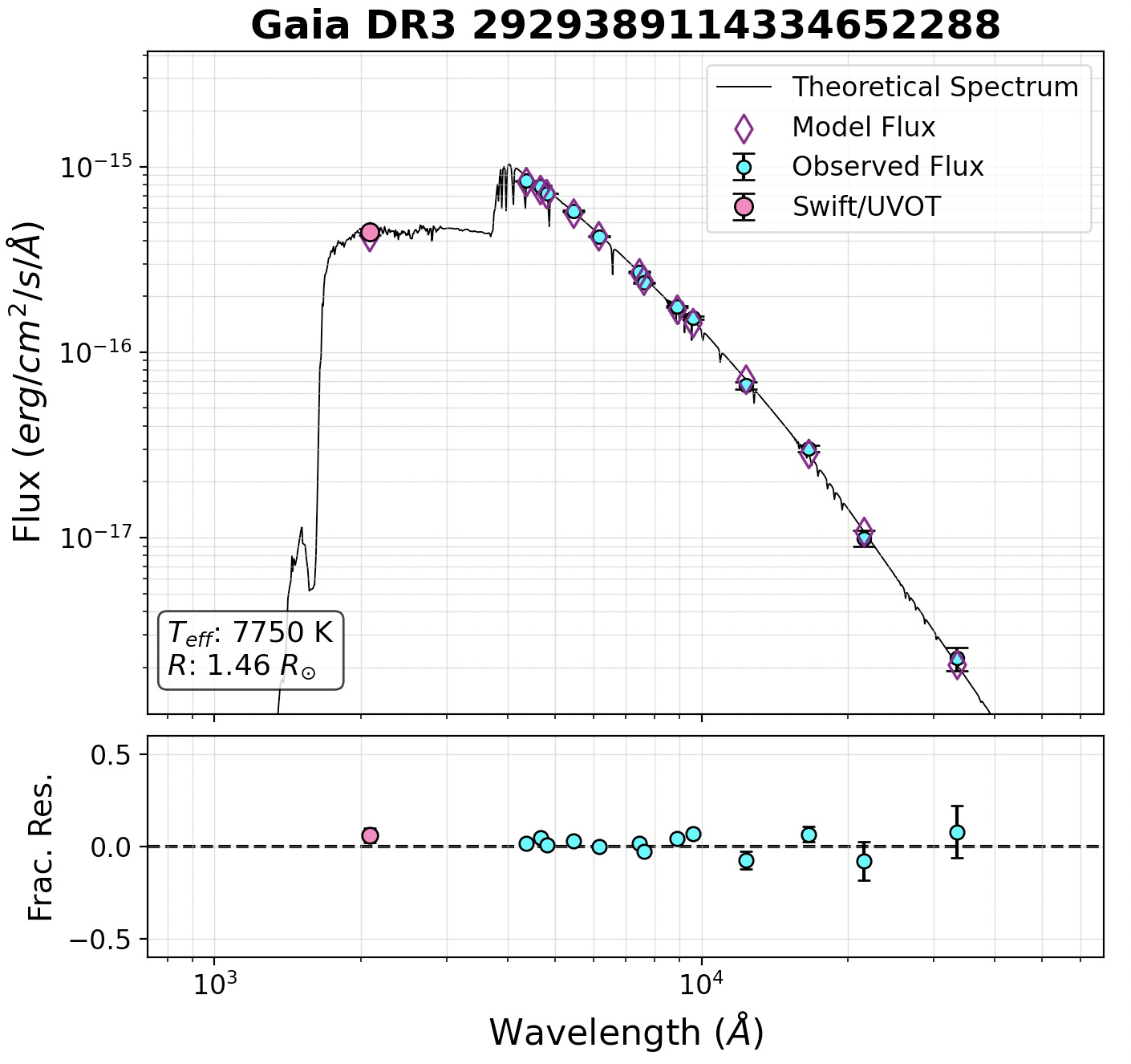}
    \caption{SED fitting for the single component BSS 16 identified as a probable member of Tombaugh~2. Top panel: Cyan circles with associated uncertainties denote the observed photometric fluxes, while the solid black curve represents the best-fitting model spectrum \citep{Castelli2003}. The \textit{Swift}/UVOT data are shown in pink circles. Purple diamonds indicate the synthetic fluxes computed for the corresponding photometric passbands. Bottom panel: Fractional residuals between the observed and model fluxes, illustrating the quality of the fit. The complete figure set (19 images) is available in the online journal.}
    \label{fig:T2_SED}
\end{figure}

To evaluate the quality of each single-component fit, we computed the fractional residuals, $(F_{\rm obs} - F_{\rm model})/F_{\rm obs}$, for all photometric points. In particular, the residuals in the \textit{Swift}/UVOT $UVM2$ and $UVW2$ bands were examined for UV excess emission. A UV excess exceeding $\sim30\%$ in at least two of these filters was taken as evidence for a potential hot companion, following the criteria of \citet{Chand2024}. This threshold minimizes false detections due to photometric uncertainties, while remaining sensitive to hot WD companions. Although chromospheric activity cannot be entirely ruled out, the observed UV excess is more consistent with a hot compact companion. However, the magnitude and persistence of the UV excess strongly favor a hot compact companion interpretation over chromospheric activity. For BSSs in Tombaugh 2, the stellar photosphere is not expected to contribute significantly in the near- and far-UV bands. Thus, a stable and measurable UV excess is most naturally explained by the presence of a hot, compact companion, most plausibly a WD. Such excesses, though occasionally arising from chromospheric activity, are most frequently attributed to WD companions in BSS systems \citep{Gosnell2015, Subramaniam2016, Rao2022}. We also searched for X-ray counterparts in archival Chandra and XMM-Newton data to probe possible coronal activity, but no detections were found for the examined stars.

\begin{figure*}
    \centering
    \includegraphics[width=0.31\linewidth]{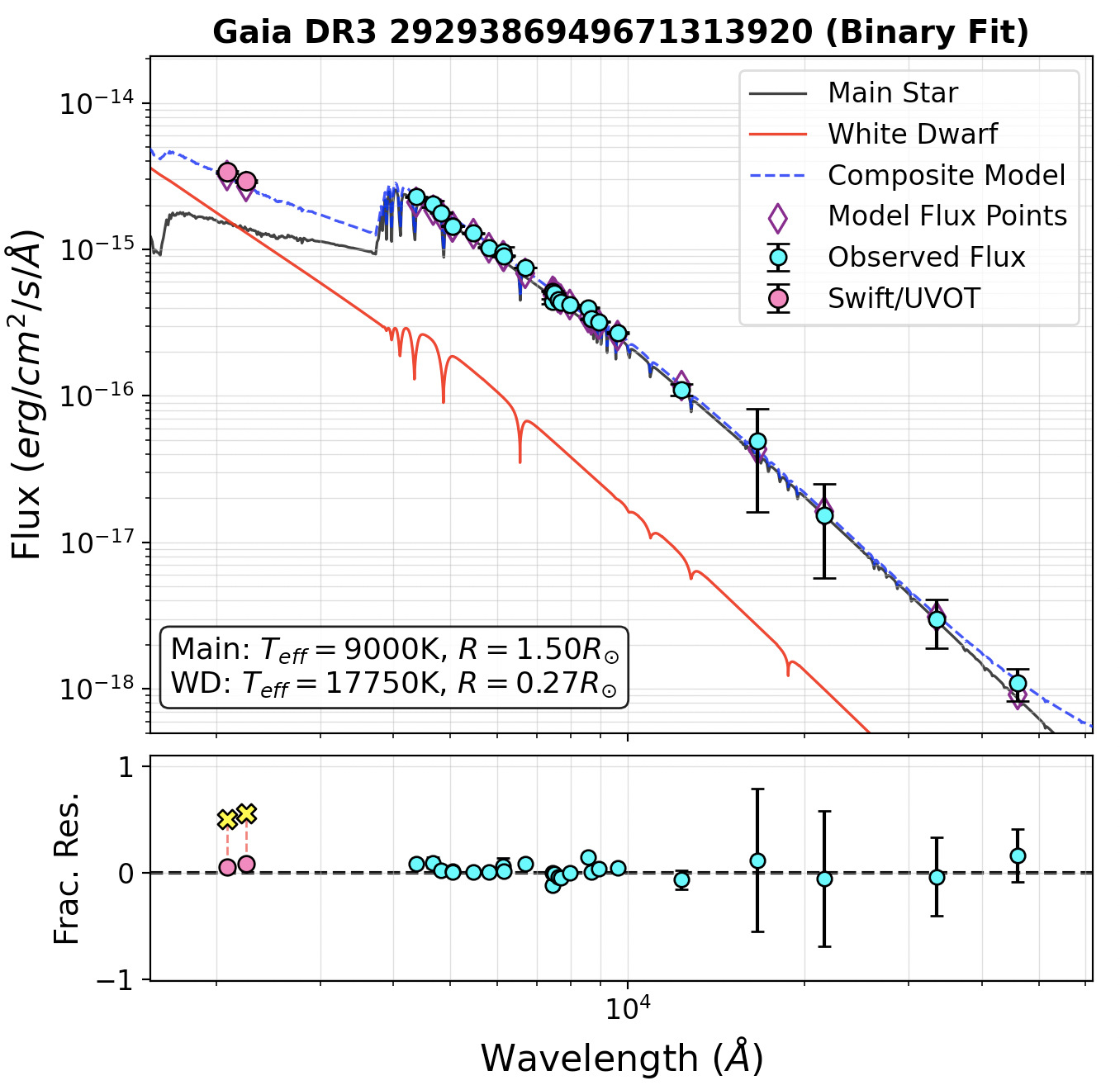}
    \includegraphics[width=0.31\linewidth]{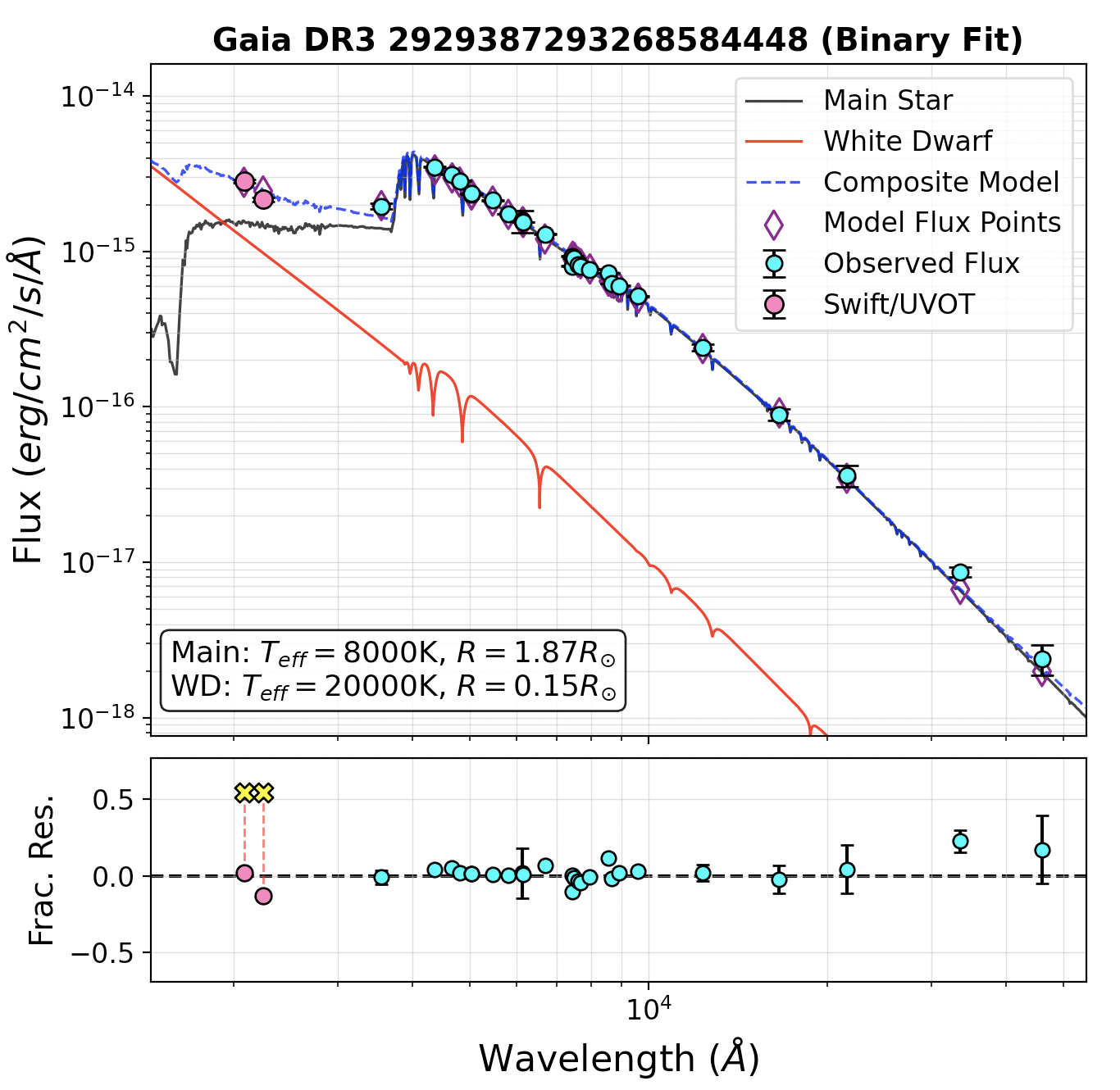}    \includegraphics[width=0.3\linewidth]{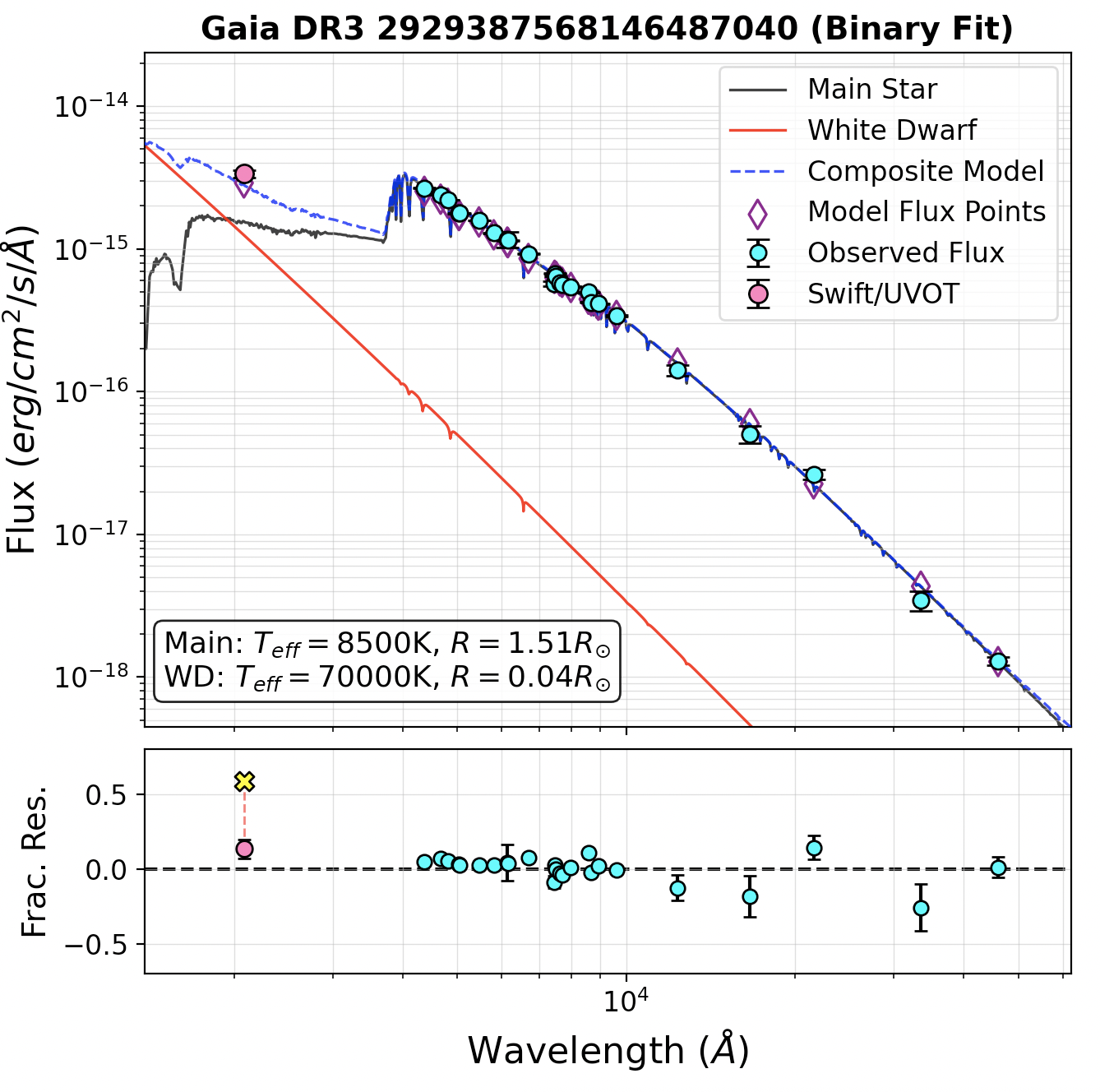}\\
    \includegraphics[width=0.31\linewidth]{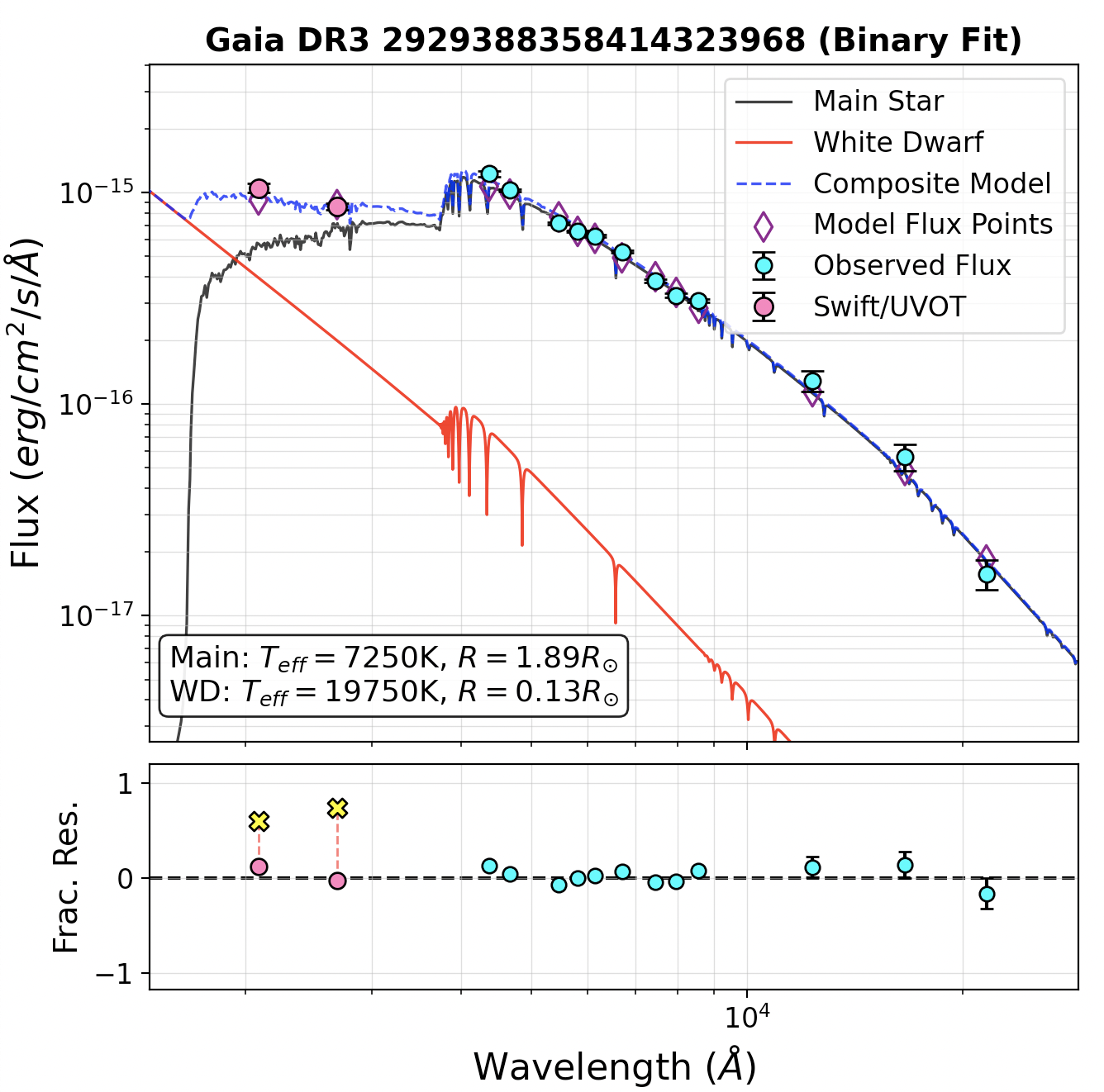}
    \includegraphics[width=0.31\linewidth]{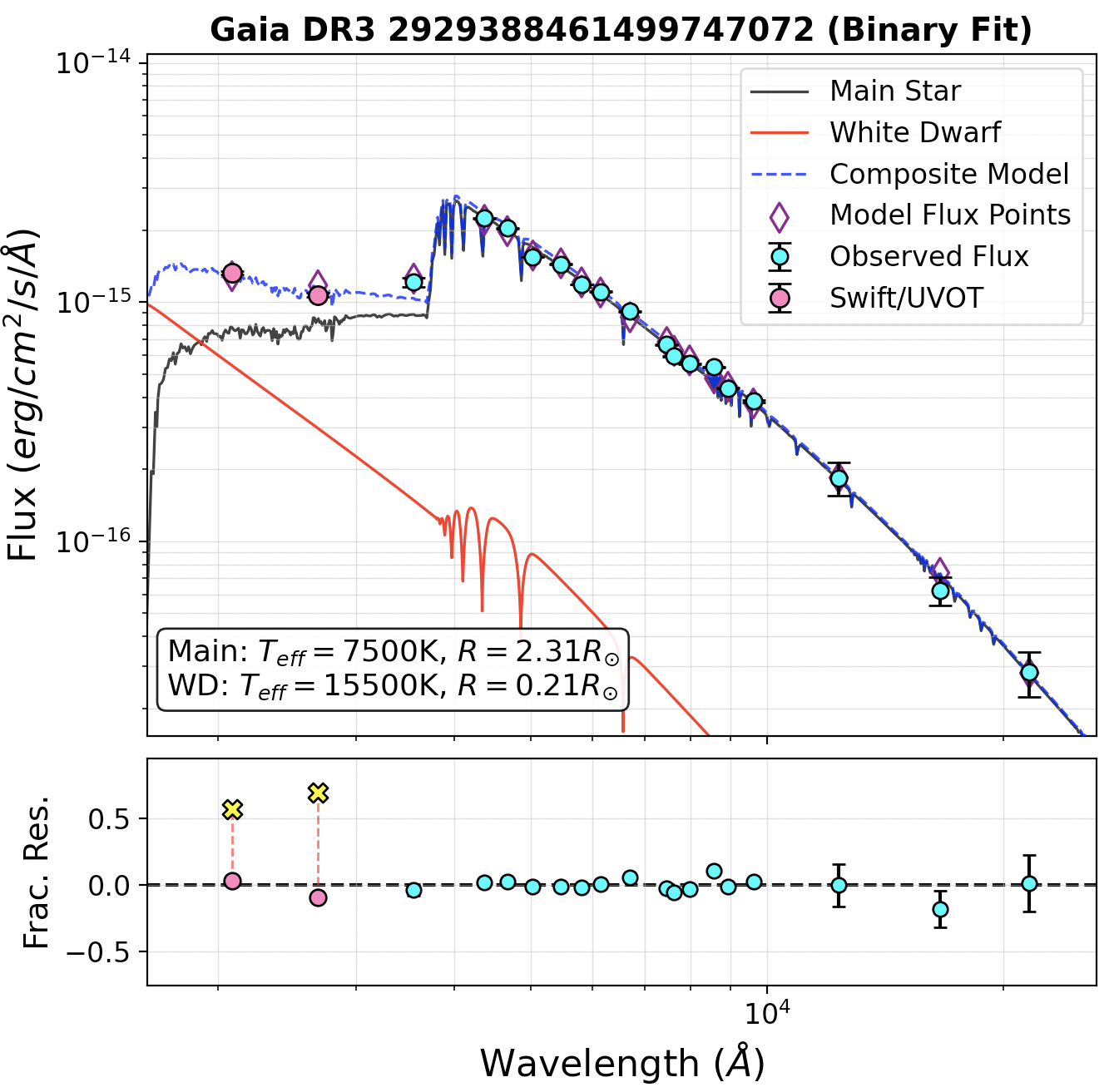}    \includegraphics[width=0.31\linewidth]{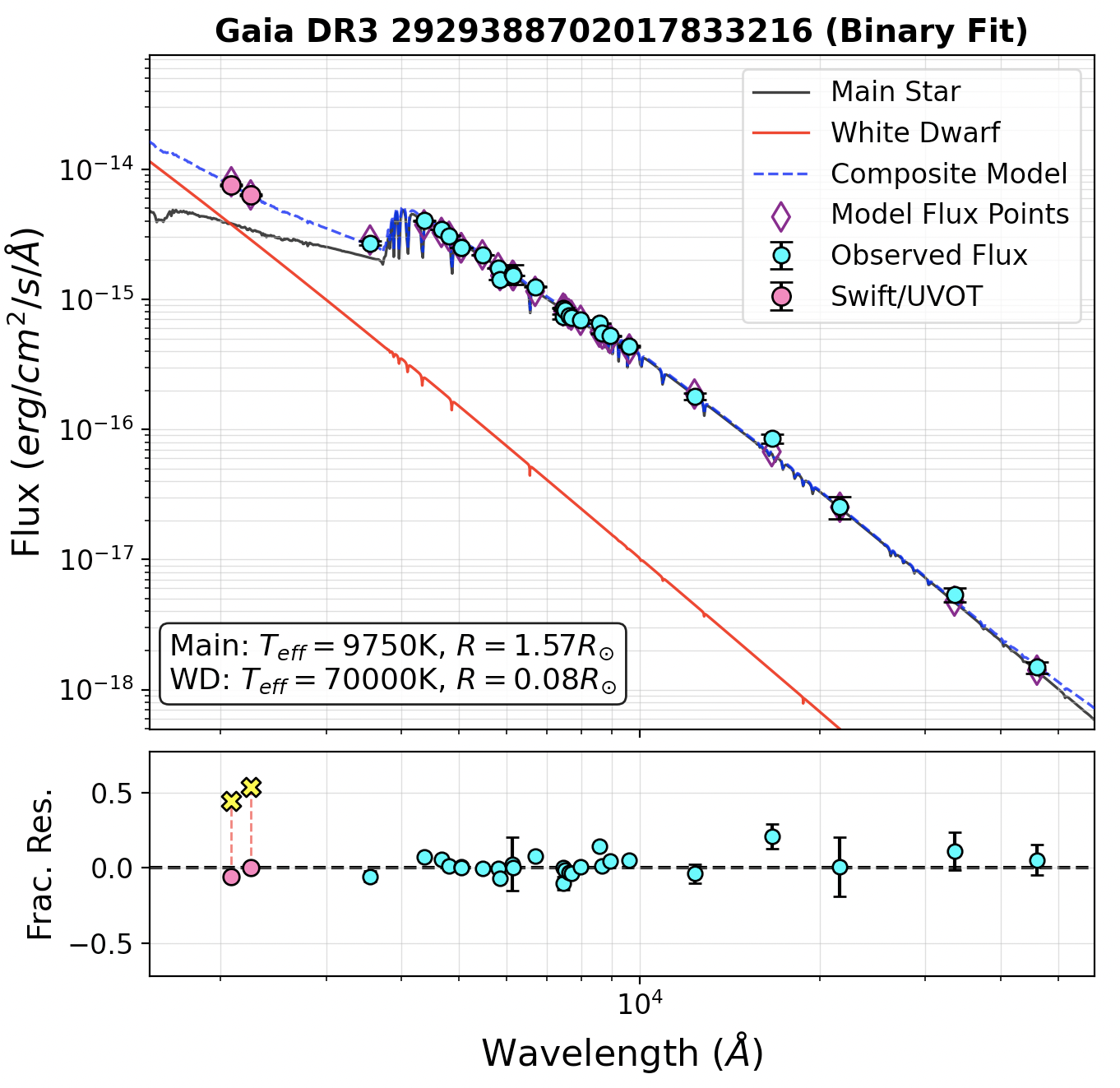}\\
    \includegraphics[width=0.31\linewidth]{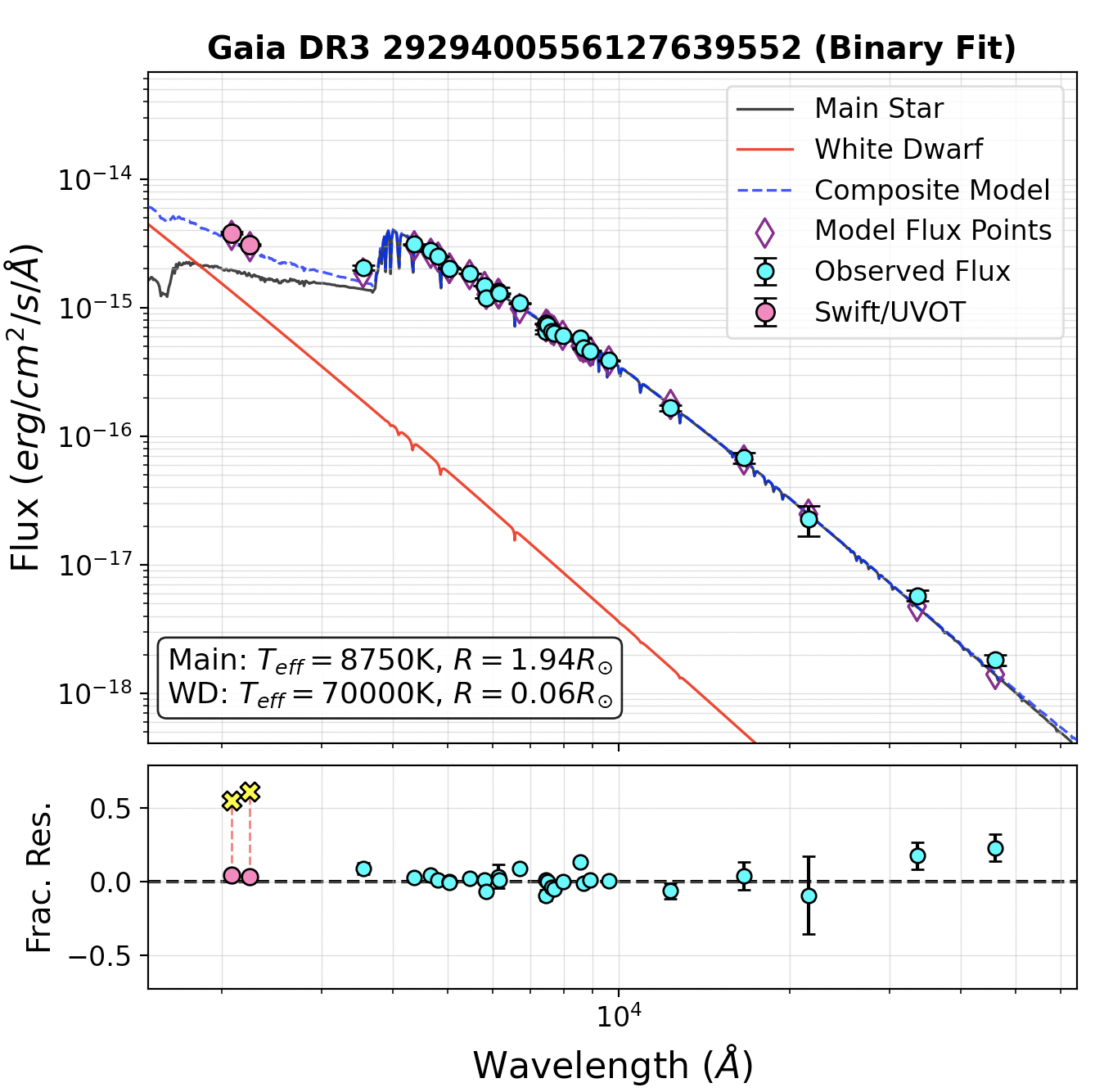}
    \includegraphics[width=0.31\linewidth]{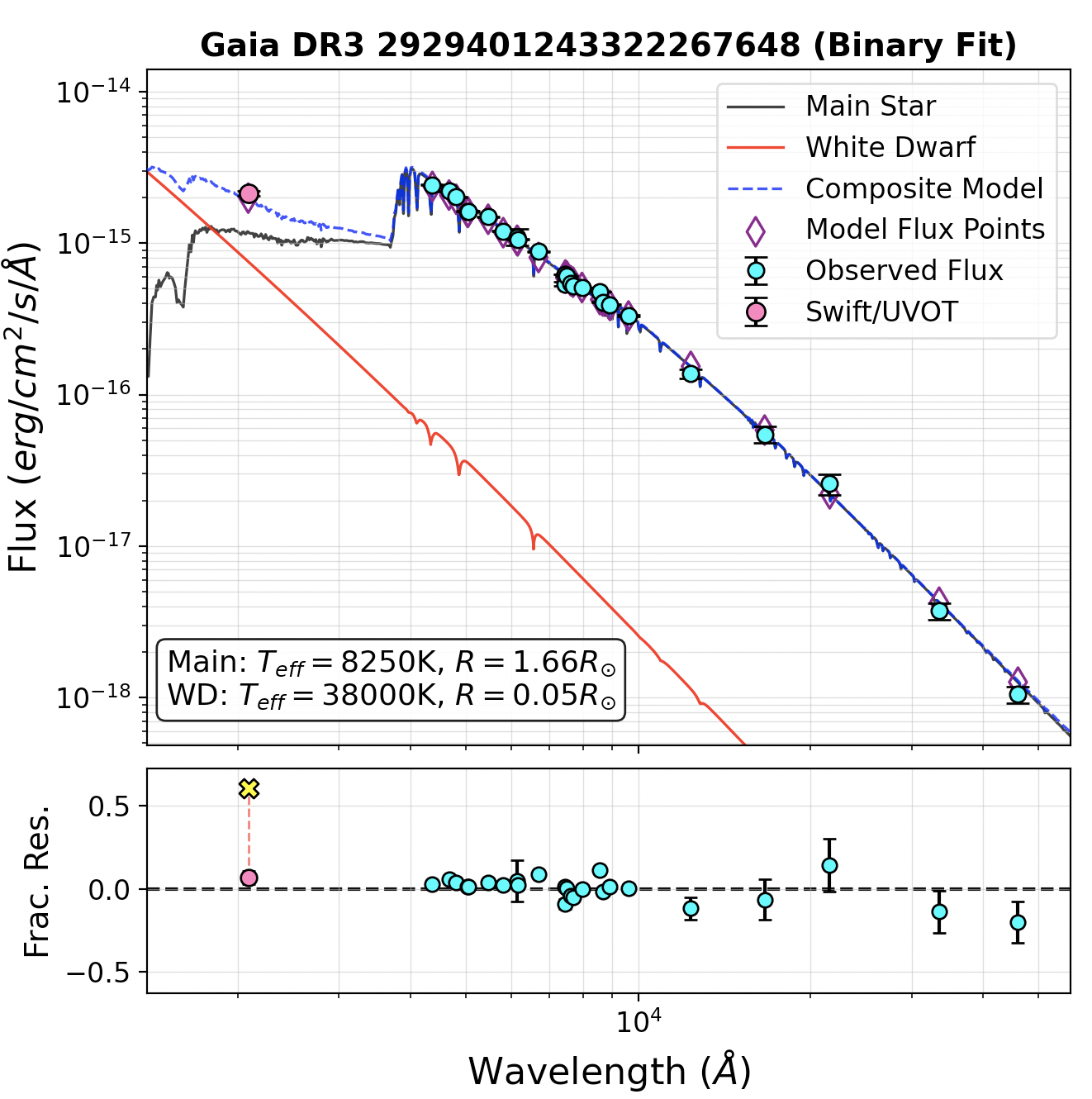}    \includegraphics[width=0.31\linewidth]{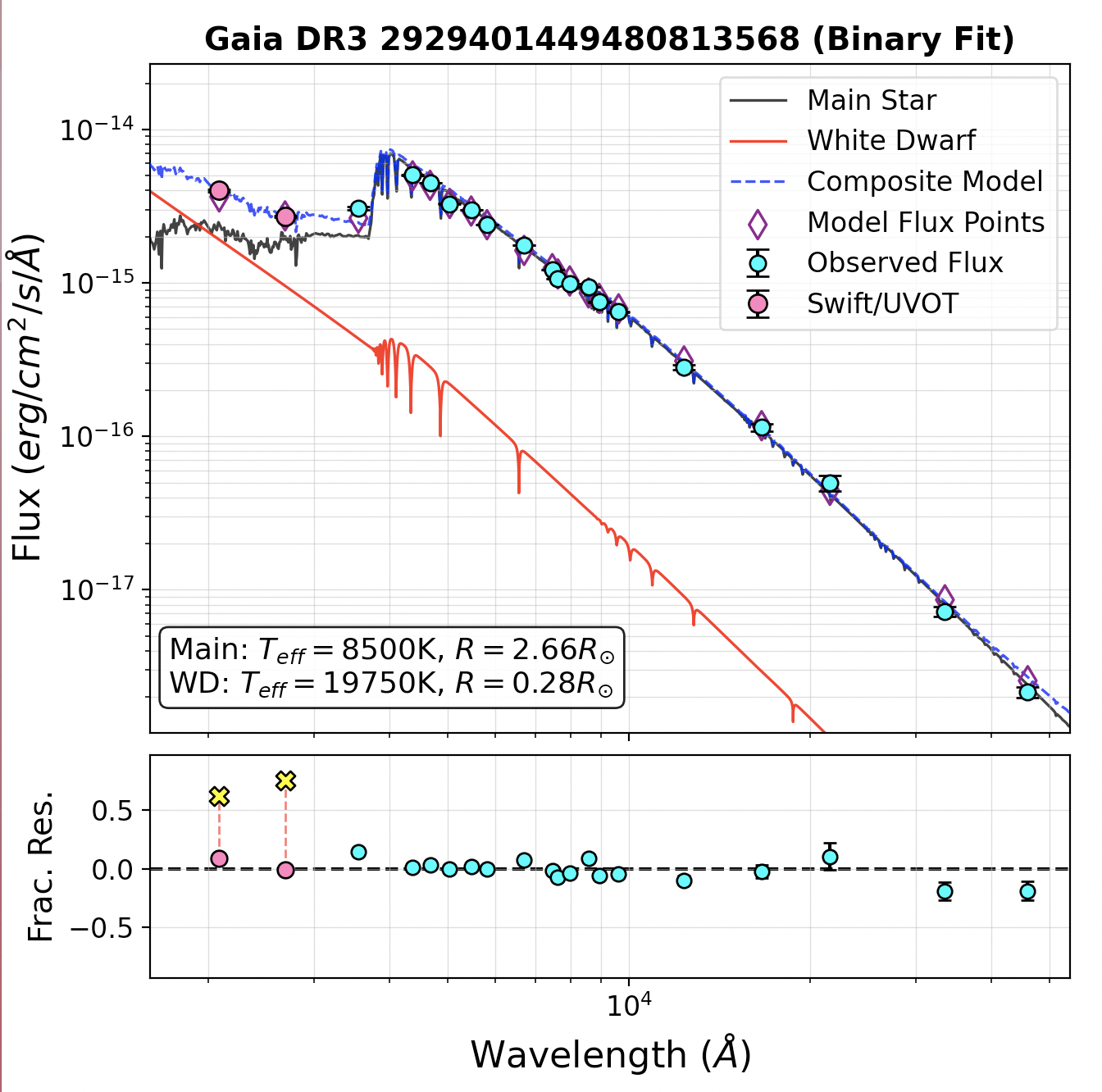}
    \caption{SED fitting analysis for binary BSS/BSS candidate systems identified as probable members of Tombaugh~2. Top panels: Observed photometric fluxes with uncertainties are shown as cyan circles, with \textit{Swift}/UVOT data in pink. The solid black and red curves correspond to the best-fitting models of \citet{Castelli2003} and \citet{Koester2010}, while the blue dashed line shows the composite spectrum. Purple diamonds denote synthetic fluxes in the relevant passbands. Bottom panels: Fractional residuals between the observed and composite model fluxes. Yellow crosses indicate residuals of the \textit{Swift}/UVOT data with respect to the single Kurucz model of the primary, highlighting the UV excess.}
    \label{fig:T2_Binary_SED}
\end{figure*}

\begin{table*}[]
\centering
\footnotesize
\caption{List of \textit{Swift}/UVOT photometric measurements ($UVW2$ and $UVM2$) for the analyzed sources.}
\label{tab:bss_yss_fluxes}
\begin{tabular}{llcccc}
\hline
No. & Gaia DR3 ID & $\alpha_{2000}$ & $\delta_{2000}$ & \multicolumn{2}{c}{UV flux (ergs s$^{-1}$ cm$^{-2}$ \AA$^{-1}$)} \\ \cline{5-6} 
                   &  & (deg) & (deg) & $UVW2$$\pm$err & $UVM2$$\pm$err \\ \hline
BSS 2  & 2929387190189411328 & 105.8411        & -20.8656        & 4.27E--16 $\pm$ 3.93E--18                                   & 6.77E--16 $\pm$ 7.48E--18 \\
BSS 3  & 2929386949671313920 & 105.7387        & -20.8475        & 1.86E--16 $\pm$ 3.42E--18                                   & 2.89E--16 $\pm$ 4.79E--18 \\
BSS 5  & 2929387293268584448 & 105.8456        & -20.8469        & 1.10E--16 $\pm$ 2.54E--18                                   & 1.51E--16 $\pm$ 4.31E--18 \\
BSS 6  & 2929387430707604480 & 105.8016        & -20.8534        & 8.91E--17 $\pm$ 2.62E--18                                   & 1.19E--16 $\pm$ 4.16E--18 \\
BSS 7  & 2929387568146487040 & 105.8224        & -20.8284        & 3.49E--17 $\pm$ 1.96E--18                                   & ---                       \\
BSS 8  & 2929387636865933696 & 105.8704        & -20.8369        & 1.10E--16 $\pm$ 2.42E--18                                   & 1.82E--16 $\pm$ 3.68E--18 \\
BSS 9  & 2929388255341328000 & 105.7771        & -20.8384        & 4.15E--16 $\pm$ 5.74E--18                                   & ---                       \\
BSS 10 & 2929388358414323968 & 105.7846        & -20.824         & 2.01E--16 $\pm$ 9.60E--18                                   & ---                       \\
BSS 11 & 2929388358421934848 & 105.784         & -20.8244        & 2.01E--16 $\pm$ 9.60E--18                                   & 1.81E--16 $\pm$ 8.15E--18 \\
BSS 12 & 2929388392780211712 & 105.7963        & -20.8127        & 3.79E--16 $\pm$ 4.53E--18                                   & 5.75E--16 $\pm$ 8.47E--18 \\
BSS 13 & 2929388461499747072 & 105.7687        & -20.8254        & 2.52E--16 $\pm$ 4.42E--18                                   & ---                       \\
BSS 15 & 2929388702017833216 & 105.8185        & -20.8094        & 2.57E--16 $\pm$ 3.54E--18                                   & 3.85E--16 $\pm$ 6.03E--18 \\
BSS 16 & 2929389114334652288 & 105.8200          & -20.7754        & 4.67E--17 $\pm$ 1.85E--18                                   & ---                       \\
BSS 17 & 2929395092929076736 & 105.8377        & -20.7392        & 6.19E--17 $\pm$ 2.56E--18                                   & 8.19E--17 $\pm$ 4.15E--18 \\
BSS 18 & 2929400556127639552 & 105.7217        & -20.7853        & 1.58E--16 $\pm$ 2.76E--18                                   & 2.29E--16 $\pm$ 4.64E--18 \\
BSS 20 & 2929393752899281920 & 105.9045        & -20.787         & 6.86E--17 $\pm$ 2.46E--18                                   & 1.16E--16 $\pm$ 5.21E--18 \\
BSS 21 & 2929398185305827328 & 105.6743        & -20.8708        & 1.02E--16 $\pm$ 3.66E--18                                   & ---                       \\
BSS 22 & 2929401243322267648 & 105.7894        & -20.7286        & 5.84E--17 $\pm$ 2.37E--18                                   & ---                       \\
BSS 23 & 2929401449480813568 & 105.6857        & -20.7489        & 4.22E--16 $\pm$ 7.00E--18                                   & ---                       \\
\hline
\end{tabular}
\end{table*}

Out of 28 analyzed sources (26 BSSs and 2 YSSs), 9 systems ($\sim$32\%) exhibit UV excess, consistent with the presence of a hot companion, yielding a sample of 19 single and 9 binary candidates. For these sources, we performed binary SED fitting using VOSA’s Binary Fit module. In this approach, the cool BSS component was modeled with a Kurucz atmosphere, while the hot companion was represented by a \citet{Koester2010} WD model. The fitting spanned $T_{\rm eff,WD} = 10,000$–$80,000$~K, with both components optimized simultaneously to minimize the global $\chi^2_\nu$ over the full UV-to-IR wavelength range. We note that the upper temperature limit of the Koester WD models (80{,}000~K) may affect the reliability of fits approaching this boundary. In particular, systems with high inferred temperatures (e.g., $T_{\mathrm{eff}} \gtrsim 70{,}000$~K) should be interpreted with caution, as the fitting procedure may be influenced by the model grid limits. 

For all binary candidates, single-component fits fail to reproduce the UV fluxes, leading to systematically larger residuals in the UV bands. The inclusion of a hot companion significantly improves the fit quality, as reflected by reduced residuals and lower reduced $\chi^2$ values ($\chi^2_\nu \approx 2$--9). On average, the binary SED fits yield an improvement of $\sim$14.5\% in $\chi^2_\nu$ compared to single-component models, indicating the statistical preference for the additional component (see Table~\ref{tab:sed_parameters_binary_wide}).

The final adopted parameters ($T_{\rm eff}$, $R$, and $L$) were taken from the best-fitting single- or binary-component models, as appropriate. The parameters listed in Table~\ref{tab:sed_parameters_single} and Table~\ref{tab:sed_parameters_binary_wide} were estimated using the statistical approach implemented in VOSA. The corresponding uncertainties were derived from a 100-iteration Monte Carlo simulation. To further validate our uncertainty estimates, particularly for hot companions approaching the model limits ($T_{\rm eff} \gtrsim 70{,}000$~K), we independently estimated the parameters and their errors using the \texttt{Binary\_SED\_Fitting} Python module \citet{2021JApA...42...89J, jadhav_2024_13928317}\footnote{\url{https://github.com/jikrant3/Binary_SED_Fitting}}. The derived reduced $\chi^2_\nu$ values and errors were highly consistent with our VOSA-based Monte Carlo approach; therefore, we retained the VOSA results to maintain methodological homogeneity across the entire sample. These values were subsequently used to interpret the stellar and binary properties of the BSS candidates. 

To assess the robustness of the SED-based classifications, we examined the potential impact of uncertainties in extinction and distance on the inferred UV excesses. Reasonable variations in these parameters primarily affect the optical and near-infrared portions of the SEDs and do not remove the excess flux observed at UV wavelengths. In all cases, the identification of UV excess, and hence the classification of BSSs as binary systems, is driven mainly by the \textit{Swift}/UVOT measurements, which provide direct sensitivity to hot companions. Thus, extinction and distance uncertainties have only a minor impact on the identification of UV excess.

\begin{table*}[]
\footnotesize
\centering
\caption{Stellar parameters derived from single-component SED fitting for BSS, BSS candidate, and YSS candidates. The table lists Gaia DR3 identifiers, effective temperatures ($T_{\rm eff}$), radii ($R$), luminosities ($L$), reduced chi-square ($\chi^2_\nu$), scaling factors applied to the SED fits, and the goodness-of-fit parameter (vgf$_b$).}
\begin{tabular}{llcccccc} \hline
No.    & Gaia DR3 ID    & $T_{\rm eff}$ (K) & $R$ ($R_\odot$) & $L$ ($L_\odot$) & $\chi^{2}_\nu$ & Scaling Factor & vgf$_b$ \\ \hline
BSS 1  & 2929385781434416256 & 8000$\pm$156      & 1.56$\pm$0.44   & 9.19$\pm$5.27   & 1.73           & 2.47E-23       & 1.55    \\
BSS 2  & 2929387190189411328 & 8250$\pm$125      & 3.67$\pm$1.45   & 55.91$\pm$46.12 & 3.94           & 2.69E-22       & 0.65    \\
BSS 4  & 2929387220244270848 & 7500$\pm$125      & 1.41$\pm$0.42   & 5.79$\pm$3.62   & 1.72           & 2.28E-23       & 0.89    \\
BSS 6  & 2929387430707604480 & 8000$\pm$125      & 1.69$\pm$0.67   & 10.84$\pm$8.91  & 3.99           & 5.63E-23       & 0.95    \\
BSS 8  & 2929387636865933696 & 8250$\pm$131      & 1.70$\pm$0.53   & 12.34$\pm$8.06  & 4.36           & 3.51E-23       & 1.31    \\
BSS 9  & 2929388255341328000 & 8000$\pm$174      & 3.53$\pm$0.99   & 53.42$\pm$31.43 & 7.82           & 1.26E-22       & 4.71    \\
BSS 11 & 2929388358421934848 & 8250$\pm$125      & 1.55$\pm$0.43   & 12.05$\pm$7.62  & 1.46           & 2.44E-23       & 2.43    \\
BSS 12 & 2929388392780211712 & 8750$\pm$125      & 2.26$\pm$0.66   & 29.45$\pm$18.08 & 2.37           & 5.58E-23       & 5.73    \\
BSS 14 & 2929388564572665216 & 8000$\pm$157      & 1.82$\pm$0.51   & 12.89$\pm$7.33  & 5.62           & 3.33E-23       & 8.85    \\
BSS 16 & 2929389114334652288 & 7750$\pm$125      & 1.46$\pm$0.40   & 7.01$\pm$4.04   & 2.21           & 1.99E-23       & 0.46    \\
BSS 17 & 2929395092929076736 & 8000$\pm$125      & 2.66$\pm$0.88   & 26.28$\pm$18.15 & 5.75           & 1.00E-22       & 0.64    \\
BSS 18 & 2929384407050539136 & 7500$\pm$125      & 1.78$\pm$0.53   & 9.10$\pm$5.78   & 2.00           & 3.57E-23       & 0.75    \\
BSS 19 & 2929384647568733696 & 7750$\pm$125      & 1.78$\pm$0.65   & 10.21$\pm$7.74  & 3.06           & 5.43E-23       & 0.59    \\
BSS 20 & 2929393752899281920 & 7500$\pm$125      & 1.69$\pm$0.49   & 8.14$\pm$5.03   & 2.50           & 3.10E-23       & 0.50     \\
BSS 21 & 2929398185305827328 & 6250$\pm$125      & 2.59$\pm$0.73   & 9.29$\pm$5.65   & 4.16           & 6.77E-23       & 0.99    \\
BSS 24 & 2929406878319329664 & 7500$\pm$139      & 2.21$\pm$0.62   & 14.04$\pm$8.70  & 5.91           & 4.96E-23       & 0.26    \\
BSS 25 & 2929407977830934400 & 7500$\pm$125      & 2.09$\pm$0.59   & 12.65$\pm$7.84  & 5.07           & 4.44E-23       & 6.49    \\
YSS 1  & 2929388530219213824 & 7000$\pm$144      & 3.26$\pm$0.92   & 22.88$\pm$14.32 & 6.92           & 1.08E-22       & 0.73    \\
YSS 2  & 2929400693566526208 & 7250$\pm$153      & 2.03$\pm$0.90   & 10.27$\pm$9.76  & 2.71           & 1.03E-22       & 0.21    \\ \hline
\end{tabular}
\label{tab:sed_parameters_single}
\end{table*}

\begin{sidewaystable}
\footnotesize
\centering
\caption{Stellar parameters derived from binary SED fitting for BSS and BSS candidate systems with hot companions. The table lists Gaia DR3 identifiers, effective temperatures ($T_{\rm eff}$), radii ($R$), and luminosities ($L$) for both the cooler primary (Component A) and the hotter secondary (Component B). The reduced chi-square ($\chi^2_\nu$) and the visual goodness-of-fit parameter (vgf$_b$) values for both single- and binary fits are provided for direct comparison, along with the individual scaling factors.}
\setlength{\tabcolsep}{5pt} 
\begin{tabularx}{\linewidth}{llcccccccccccc} \hline
\multirow{2}{*}{No.} & \multirow{2}{*}{Gaia DR3 ID} & \multicolumn{3}{c}{Component A (Cooler)} & \multicolumn{3}{c}{Component B (Hotter)} & \multicolumn{2}{c}{$\chi^{2}_\nu$} & \multicolumn{2}{c}{Scaling Factor} & \multicolumn{2}{c}{vgf$_b$} \\ \cline{3-5} \cline{6-8} \cline{9-10} \cline{11-12} \cline{13-14}
 &  & $T_{\rm eff}$ (K) & $R$ ($R_\odot$) & $L$ ($L_\odot$) & $T_{\rm eff}$ (K) & $R$ ($R_\odot$) & $L$ ($L_\odot$) & Single & Binary & Comp A & Comp B & Single & Binary \\ \hline
BSS 3  & 2929386949671313920 & 9000$\pm$860 & 1.50$\pm$0.44 & 13.47$\pm$7.89 & 17750$\pm$6360 & 0.27$\pm$0.08 & 6.61$\pm$3.86 & 5.33 & 4.45 & 2.44E-23 & 7.85E-25 & 3.00 & 2.27 \\
BSS 5  & 2929387293268584448 & 8000$\pm$125 & 1.87$\pm$0.70 & 12.99$\pm$9.74 & 20000$\pm$1580 & 0.15$\pm$0.06 & 3.29$\pm$2.46 & 5.57 & 4.78 & 6.20E-23 & 4.05E-25 & 1.84 & 1.42 \\
BSS 7  & 2929387568146487040 & 8500$\pm$125 & 1.51$\pm$0.55 & 10.94$\pm$7.98 & 70000$\pm$5477 & 0.04$\pm$0.02 & 7.01$\pm$5.10 & 4.14 & 3.74 & 3.81E-23 & 3.34E-26 & 1.58 & 1.44 \\
BSS 10 & 2929388358414323968 & 7250$\pm$125 & 1.89$\pm$0.53 & 9.70$\pm$5.52 & 19750$\pm$3500 & 0.13$\pm$0.04 & 2.46$\pm$1.39 & 2.59 & 2.17 & 3.61E-23 & 1.76E-25 & 9.49 & 8.10 \\
BSS 13 & 2929388461499747072 & 7500$\pm$125 & 2.31$\pm$0.65 & 15.17$\pm$8.63 & 15500$\pm$3700 & 0.21$\pm$0.06 & 2.39$\pm$1.35 & 14.12 & 13.92 & 5.37E-23 & 4.66E-25 & 1.05 & 0.92 \\
BSS 15 & 2929388702017833216 & 9750$\pm$125 & 1.57$\pm$0.54 & 20.10$\pm$13.75 & 70000$\pm$3200 & 0.08$\pm$0.03 & 24.59$\pm$16.79 & 6.77 & 5.82 & 3.65E-23 & 9.69E-26 & 2.04 & 1.06 \\
BSS 18 & 2929400556127639552 & 8750$\pm$125 & 1.94$\pm$0.57 & 19.97$\pm$11.67 & 70000$\pm$5000 & 0.06$\pm$0.02 & 13.00$\pm$7.58 & 5.40 & 4.68 & 4.06E-23 & 4.109E-26 & 0.98 & 0.87 \\
BSS 22 & 2929401243322267648 & 8250$\pm$125 & 1.66$\pm$0.55 & 11.62$\pm$7.75 & 38000$\pm$9300 & 0.05$\pm$0.02 & 3.93$\pm$2.61 & 4.94 & 3.59 & 3.88E-23 & 4.22E-26 & 0.93 & 0.78 \\
BSS 23 & 2929401449480813568 & 8500$\pm$125 & 2.65$\pm$0.37 & 34.17$\pm$9.82 & 19750$\pm$3650 & 0.28$\pm$0.04 & 11.48$\pm$3.24 & 10.63 & 8.76 & 7.14E-23 & 8.13E-25 &2 .92 & 2.43 \\
\hline
\end{tabularx}
\label{tab:sed_parameters_binary_wide}
\end{sidewaystable}

A comparison of the results from \citet{Sheikh2024} and \citet{Zeng2025} highlights how effective the joint use of \textit{Gaia} DR3 astrometry, broad-band photometry, and VOSA-based SED fitting can be for studying BSSs in OCs. Both groups relied on machine-learning methods for membership selection. They performed SED analyses from the UV to the IR, reaching a consistent conclusion that binary mass transfer is the dominant pathway for BSS formation. While the number of BSSs in Tombaugh 2 (19) differs from those in NGC 2243 (12) and NGC 6134 (3), which may reflect intrinsic cluster differences, we note that with only two clusters, it is difficult to generalize, and UV observations primarily detect recently formed BSSs due to the cooling of WD companions over time.  Therefore, the inferred binary fraction based on UV detections should be considered a lower limit. Quantitative discrepancies, such as differences in mass estimates or metallicity, likely arise from choices of isochrones (MIST vs. PARSEC), machine-learning frameworks, and auxiliary datasets (e.g., \textit{Swift}/UVOT and Pan-STARRS vs. \textit{NEOWISE} and \textit{TESS}).

As a result of this analysis, the sample comprises 19 single and nine binary systems. The SED fits were carried out under the single-star assumption using the \citet{Castelli2003} atmosphere models; representative single-component fits are shown in Figure~\ref{fig:T2_SED}. Figure~\ref{fig:T2_Binary_SED} shows the SED fits for the binary candidates, while the full set of figures is available as Figure Set 1. Most stars that are well fit with single-component models have $T_{\rm eff} \simeq 7000$–$9000$~K, radii of $R \simeq 1.4$–$2.6\,R_{\odot}$, and luminosities around $6$–$15\,L_{\odot}$. Their reduced chi-square values ($\chi^2_r \simeq 1.5$–$9$) indicate generally reliable fits. Cooler single stars with $T_{\rm eff} \approx 6000$–$6500$~K exhibit somewhat larger radii and luminosities of about $8$–$10\,L_{\odot}$, consistent with subgiant evolution. The binary systems constitute the most distinctive subgroup in Tombaugh~2. The primaries (Component~A) lie close to the cluster TO or on the subgiant branch, with effective temperatures of $T_{\rm eff}\simeq 7500$–$9000$~K, radii of $1.5$–$2.3\,R_{\odot}$, and luminosities of roughly $10$–$20\,L_{\odot}$. The SED analysis indicates that the hot secondary components (Component~B) separate naturally into two evolutionary classes, defined primarily by their temperatures and radii. For clarity, we divide the hot companions into two empirical groups as follows:

\begin{figure*}[ht]
\centering
\includegraphics[width=0.85\linewidth]{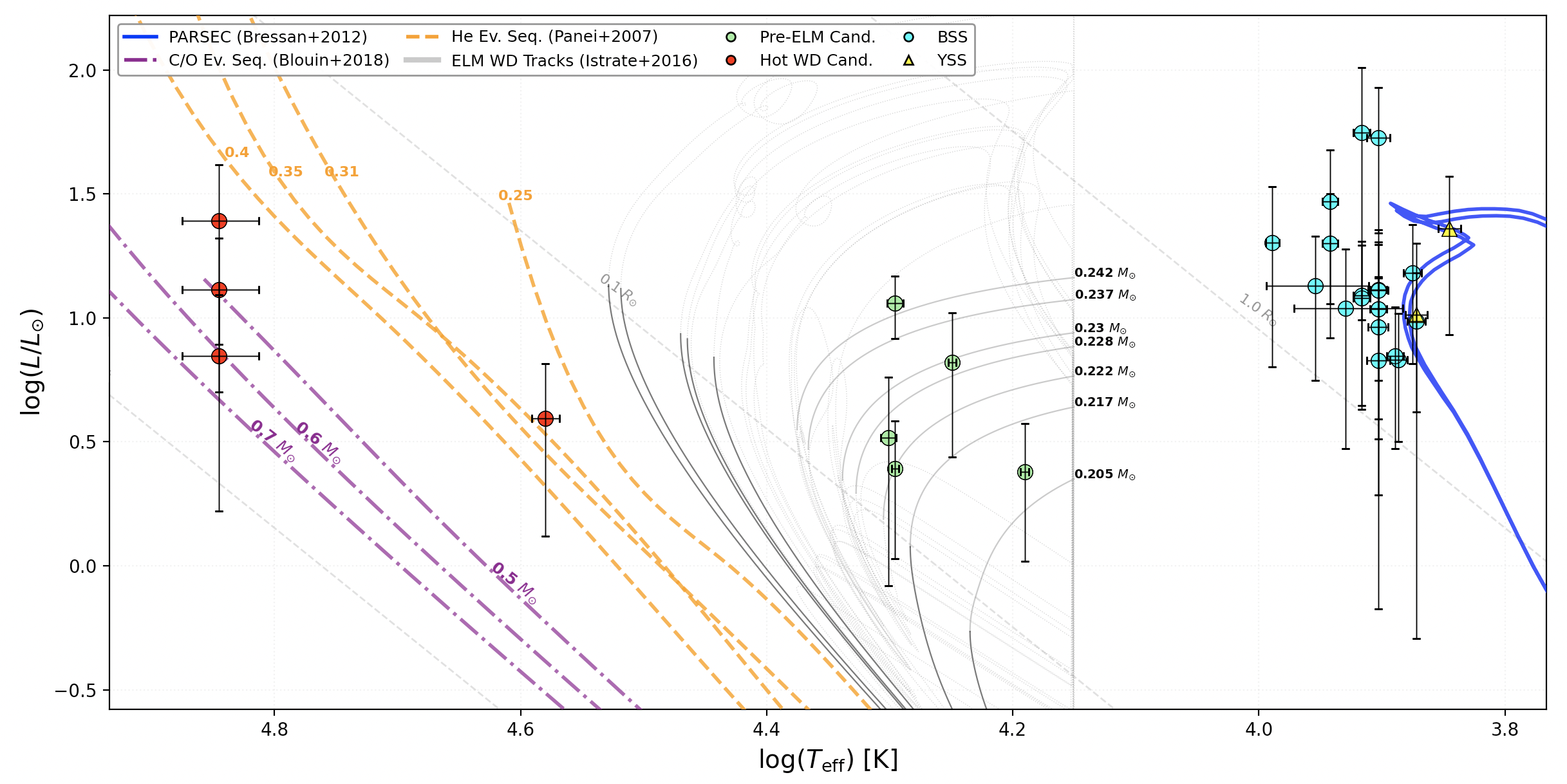}
\caption{HR diagram illustrating the evolutionary status of the hot secondary components. The gray curves trace the evolutionary tracks of He-core pre--WDs with masses between $0.20$ and $0.25\,M_{\odot}$ \citep{Istrate2016}. The solid black segments correspond to the quiescent cooling phase, while the dotted segments indicate evolutionary loops associated with hydrogen-shell flashes. The magenta dash-dotted line shows the evolutionary sequence of C/O-core WDs \citep{Blouin2018}. The dashed orange curves represent the cooling sequences of low-mass He-core WDs with masses ranging from $0.25$ to $0.4\,M_{\odot}$ \citep{Panei2007}. The solid blue line corresponds to the PARSEC \citep{Bressan2012} isochrones adopted for the host cluster. Dashed gray lines indicate loci of constant stellar radius for reference.}
    \label{fig:HR_diagram}
\end{figure*}

\textit{(i) Group~I: Pre-ELM WD candidates:} The first group consists of systems with effective temperatures in the range  $T_{\rm eff}\simeq 15{,}500$--$20{,}000$~K and inferred radii of $R\simeq 0.13$--$0.28\,R_{\odot}$ (e.g. systems BSS 3, BSS 5, BSS 10, BSS 13, BSS 23). These radii are significantly larger than those expected for fully degenerate carbon--oxygen WDs ($R\sim 0.01\,R_{\odot}$), but are consistent with the properties of \emph{proto--WDs} or the precursors of \emph{extremely low-mass (ELM) WDs} formed through binary mass stripping. The inferred temperatures and radii of the hot companions correspond to post--mass-transfer evolutionary phases with characteristic lifetimes of $\sim10^{7}$--$10^{8}$~yr, which are easily accommodated within the $\sim1.74$~Gyr age of Tombaugh~2. This confirms that the detected hot components are physically plausible and are recent remnants of binary mass transfer. Binary evolution models predict that low-mass ($\lesssim 0.3\,M_{\odot}$) Helium-core remnants can retain extended, hydrogen-rich envelopes for a substantial fraction of their post-mass-transfer evolution, appearing as inflated objects at relatively high effective temperatures before contracting onto the final WD cooling sequence \citep{Istrate2016}.

To validate this evolutionary status, we positioned the Group~I candidates on the Hertzsprung-Russell (HR) diagram alongside the evolutionary tracks for low-mass helium-core pre-WDs (Figure~\ref{fig:HR_diagram}). The candidates align remarkably well with the theoretical ``knee'' of the contraction phase for masses between 0.20 and 0.25M$_\odot$. Although the luminosity uncertainties are significant, a natural consequence of disentangling the SED of a faint companion from the primary star that dominates the optical flux, the positions of these targets are consistent with the theoretical tracks at the 1$\sigma$ confidence level, strongly supporting their classification as contracting Pre-ELM objects. Observational support for this evolutionary stage is provided by pulsating pre-He WDs identified in compact binaries  \citep{Cakirli2025}. The radii inferred for our systems are comparable to those of well-studied bloated proto--WDs, such as the stripped companion in 
WASP~0247--25 ($T_{\rm eff}\approx 17{,}000$~K, $R\approx 0.24\,R_{\odot}$; \citet{Maxted2013}). Given their elevated temperatures, inflated radii, and binary nature, a main-sequence or subgiant interpretation is unlikely. We therefore interpret these objects as stripped stellar cores currently undergoing gravitational contraction toward the ELM WD cooling track.

\textit{(ii) Group~II: Young hot WDs}
The second group comprises substantially hotter ($T_{\rm eff}\approx 38{,}000$--$70{,}000$~K) and more compact objects, with inferred radii in the range $R\simeq 0.04$--$0.08\,R_{\odot}$. While fully cooled WDs typically have radii of order $\sim 0.01\,R_{\odot}$, the larger values derived here are consistent with \emph{young WDs} that have only recently entered the cooling track. At such high effective temperatures, particularly for $T_{\rm eff} \gtrsim 70{,}000$~K, the stellar structure is supported not only by electron degeneracy pressure, but also by residual thermal pressure in the outer, partially non-degenerate layers \citep{Althaus2013}. As a result, these hot remnants appear inflated relative to older, cooler WDs. In the HR diagram (Figure~\ref{fig:HR_diagram}), the Group~II objects occupy the high-temperature regime ($T_{\rm eff}\gtrsim 30{,}000$~K), clearly separated from the Pre-ELM candidates. Notably, the candidate with $T_{\rm eff} \approx 38{,}000$~K and $\log(L/L_{\odot}) \approx 0.59$ aligns exceptionally well with the cooling sequences of low-mass He-core WDs in the $0.31\,M_{\odot}$ \citep{Panei2007}. The remaining hotter candidates overlap with the early cooling stages of C/O-core WD evolutionary sequences. By comparing their positions in the HR diagram with the C/O-core cooling tracks\footnote{\url{astro.umontreal.ca/~bergeron/CoolingModels/}} of \citet{Blouin2018}, we estimate the masses of these hot WD candidates to be in the range $\sim 0.5$--$0.7\,M_{\odot}$. Unlike the Group~I systems, which trace the nearly constant-luminosity contraction phase of He-core pre--WDs, the Group~II candidates are consistent with remnants that have already settled onto, or are approaching, the canonical WD cooling sequence. We emphasize that the division into "Pre-ELM" (Group I) and "young hot WDs" (Group II) is based on broad-band SEDs and HR-diagram locations. Spectroscopic confirmation via Balmer line profiles and surface-gravity measurements would be required for a definitive classification of the hot companions. Nevertheless, they represent stripped remnants produced by binary mass transfer, and the evolutionary distinction does not affect our primary conclusion that mass transfer is the dominant formation channel of BSSs in Tombaugh~2.

In interpreting the SED fits for systems that may host hot compact companions, it is useful to consider the relative flux contributions across different wavelength ranges. Because these companions can contribute only a small fraction of the optical flux while dominating the UV emission, the corresponding SED fits may yield comparatively large reduced chi-square values ($\chi_{r}^{2}\approx 2$--14), particularly in systems with strong UV--optical contrast. Elevated $\mathrm{vgf}_{b}$ values observed in some binaries may also indicate that modest photometric mismatches or limitations in the adopted atmosphere models become more noticeable when a faint, hot secondary component is present.

Tombaugh~2 is an intermediate-age open cluster with slightly sub-solar metallicity ([Fe/H] = -0.35 $\pm$ 0.04 dex), and has been less frequently studied in the context of BSS with hot companions. Well-studied clusters such as M67 and NGC 188 have provided important reference cases for BSS formation, while Tombaugh 2 represents a different combination of age and chemical composition \citep{2016ApJ...833L..27S, 2018MNRAS.481..226S}. The presence of BSS candidates with UV excesses consistent with hot, compact companions indicates that binary evolution involving mass transfer and white-dwarf remnants can also occur in clusters like Tombaugh 2. Our results, therefore, add Tombaugh 2 to the small number of OCs in which BSS–WD systems have been identified.

The identification of UV excesses and hot compact companions through SED decomposition provides strong evidence for past or ongoing mass transfer in a significant fraction of the BSS population. However, SED-based binarity alone cannot constrain orbital properties or confirm dynamical interaction. For this reason, we complement the SED analysis with multi-epoch RV measurements, which offer an independent and direct diagnostic of binarity. In the following section, we examine whether the BSSs exhibiting UV excesses and stripped companions also exhibit RV variability consistent with that of interacting or post-interaction binary systems.

\section{Radial Velocity Analysis and Binary Status for BSS} \label{sec:RV_analysis}

A subset of the identified BSS candidates has multi-epoch spectroscopic observations from the VLT/FLAMES archive, with 11 of the 26 BSS stars having data that allow us to assess RV variability among a significant subset of the population. These stars are highlighted in Figure \ref{fig:cmd} and form the basis of the RV analysis presented in Section~\ref{sec:RV_analysis}, allowing us to connect their photometric and SED-based properties with direct kinematic evidence for binarity. The primary aim of the RV  analysis is not to derive detailed orbital solutions, but to test whether BSSs identified as binary systems through UV SED decomposition exhibit statistically significant RV variability. RVs provide an independent dynamical diagnostic of both cluster membership and binarity, complementing the photometric and SED-based evidence for mass transfer. In this context, RV variability serves as a direct kinematic proxy for detecting hot stripped companions, allowing us to assess whether these systems are consistent with interacting or post-interaction binaries. We derived heliocentric RVs using the \textsc{iSpec} spectroscopic toolkit \citep{Blanco-Cuaresma2014}, cross-correlating the observed spectra with appropriate synthetic templates. The availability of multiple observing epochs (typically 3–5 exposures per star) spanning approximately one month enables us to probe intrinsic RV variability in the BSS candidates, rather than relying solely on mean velocities. Given the limited number of spectroscopic epochs and the temporal sampling of the available data, a detailed orbital solution or precise derivation of binary parameters is not feasible.

The derived RV statistics are presented in Table \ref{tab:rv_stats}. For each target, the mean RV ($\langle V_{r} \rangle$) is reported together with the internal measurement uncertainty from iSpec, while the standard deviation of the individual epochs ($\sigma_{V_{r}}$) is provided as a measure of the observed variability. This presentation allows a direct comparison between the internal errors and the observed scatter. The latter is used as a proxy for RV variability to identify binary candidates.

We emphasize that the observed RV dispersions ($\sigma_{V_r}$) are significantly larger than the internal uncertainties ($\epsilon_{V_r}$), indicating that the detected variability is intrinsic rather than arising from measurement errors. This provides independent kinematic evidence supporting the presence of binary companions, consistent with the SED-based identification of hot stripped stars in interacting or post-interaction systems. We note that BSSs formed through merger channels are not expected to exhibit significant short-term RV variability. Therefore, the detected RV variations in several systems strongly support a binary mass-transfer origin for these objects.

\begin{figure}
    \centering
    \includegraphics[width=1\linewidth]{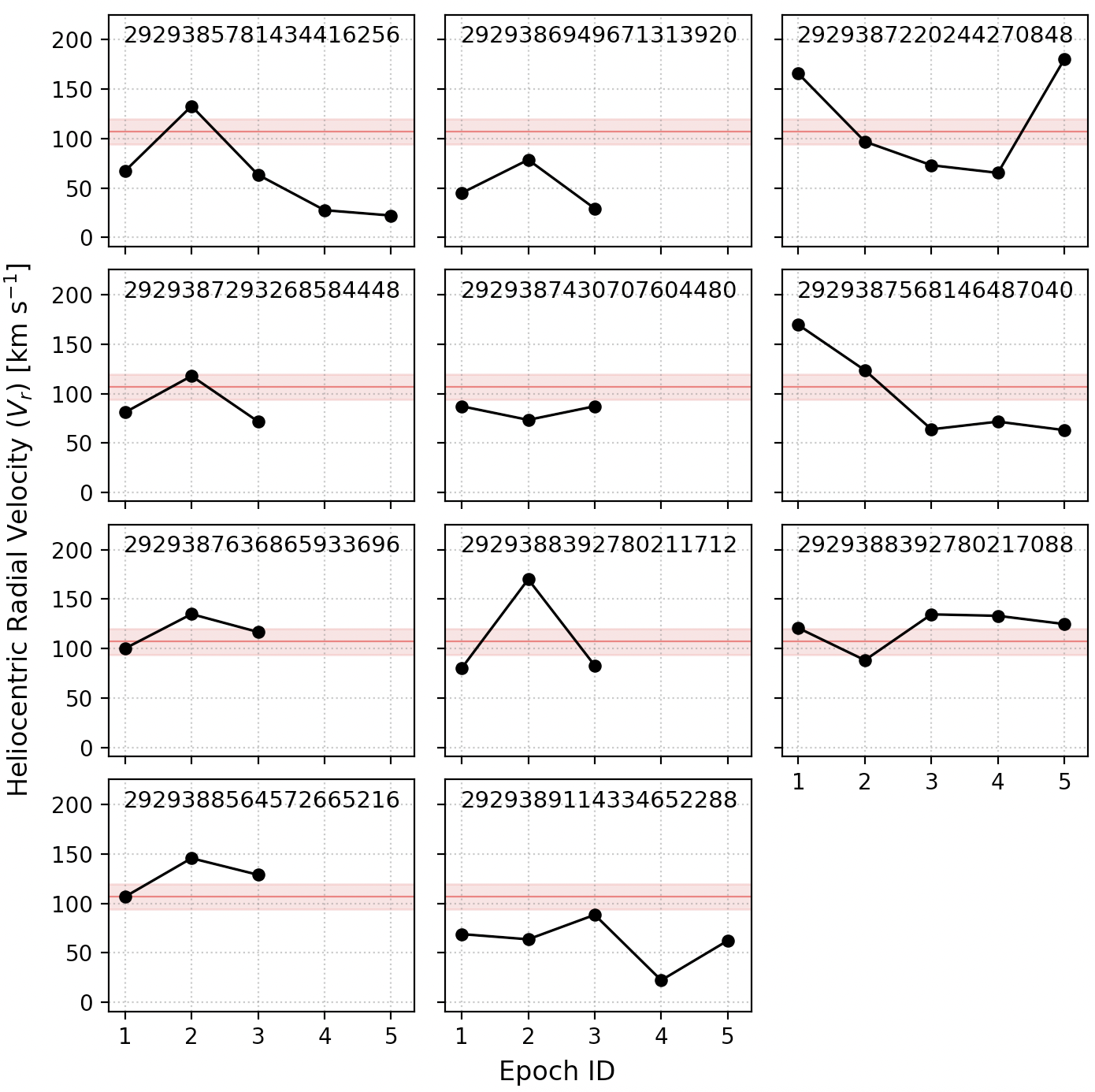}
    \caption{RV variations of the BSS candidates across five observing epochs. The horizontal axis corresponds to the Epoch IDs listed in Table \ref{tab:obs_log}. Solid circles connected by lines represent the individual heliocentric RV measurements. The solid red line indicates the determined systemic velocity of Tombaugh 2, while the red shaded region marks the $\pm 1\sigma$ velocity dispersion of the cluster members.}
\label{fig:rv_variations}
\end{figure}

We adopted the systemic velocity of Tombaugh 2 as $\langle V_{r} \rangle_{\text{cluster}} = 106.9 \pm 12.9$ km s$^{-1}$. To visualize the kinematic behavior of our sample relative to the cluster, we plotted the individual RV measurements against the observing epoch IDs in Figure \ref{fig:rv_variations}. In this figure, the solid red line and the shaded region represent the cluster's systemic velocity and the $\pm 1\sigma$ dispersion, respectively. A closer inspection of the individual epochs reveals significant RV variations ($\Delta V_{r}$) in the majority of targets, a hallmark of binary systems with relatively short periods.

\begin{table}
\centering
\footnotesize
\caption{RV measurements for BSS candidates, with uncertainties from iSpec. Columns list the Gaia DR3 source identifier, the mean heliocentric RV with internal uncertainty, the observed $V_{r}$ dispersion, and the number of observations.}
\label{tab:rv_stats}
\begin{tabular}{l l c c c}
\hline \hline
No. & Gaia DR3 Source ID & $\langle V_{r} \rangle$& $\sigma_{V_r}$ & $N_{\mathrm{obs}}$ \\
\cline{3-4}
    &               & \multicolumn{2}{c}{(km s$^{-1}$)}           & \\ 
\hline
BSS 1  & 2929385781434416256 & ~49.40 $\pm$ 0.31 & 44.22 & 5 \\
BSS 3  & 2929386949671313920 & ~47.73 $\pm$ 0.56 & 25.19 & 3 \\
BSS 4  & 2929387220244270848 & 115.62 $\pm$ 0.33 & 53.40 & 5 \\
BSS 5  & 2929387293268584448 & ~91.71 $\pm$ 0.57 & 24.37 & 3 \\
BSS 6  & 2929387430707604480 & ~83.30 $\pm$ 1.09 & ~7.96  & 3 \\
BSS 7  & 2929387568146487040 & ~92.49 $\pm$ 0.34 & 47.07 & 5 \\
BSS 8  & 2929387636865933696 & 114.24 $\pm$ 0.56 & 17.39 & 3 \\
BSS 12 & 2929388392780211712 & 112.14 $\pm$ 0.85 & 51.16 & 3 \\
BSS 14 & 2929388564572665216 & 136.57 $\pm$ 0.41 & 19.54 & 3 \\
BSS 16 & 2929389114334652288 & ~64.69 $\pm$ 0.15 & 24.12 & 5 \\
BSS 19 & 2929388392780217088 & 113.23 $\pm$ 0.31 & 18.82 & 5 \\
\hline
\end{tabular}
\end{table}

Specifically, as shown in the top-right panel of Figure \ref{fig:rv_variations}, the star BSS 4 exhibits a mean velocity ($115.62$ km s$^{-1}$) consistent with the cluster rest frame, yet its individual measurements fluctuate dramatically between $65$ and $180$ km s$^{-1}$ within a few weeks. Similarly, BSS 1 and BSS 7 display peak-to-peak velocity variations exceeding $100$ km s$^{-1}$. The large excursions of these data points well beyond the cluster's velocity dispersion region (red shaded area in Figure \ref{fig:rv_variations}) cannot be attributed to pulsation or atmospheric jitter and are conclusive evidence of binarity.

Other targets, such as BSS 12 and BSS 16, also show clear variability ($\Delta V_{r} > 40$ km s$^{-1}$), confirming them as binary candidates. In fact, BSSs analyzed in our sample exhibit RV scatter that is significantly larger than the individual measurement errors ($\sigma_{obs} \gg \langle \epsilon \rangle$), with the exception of BSS 6, which does not show significant RV variability. In fact, 10 out of the 26 BSSs in our sample exhibit RV variability or other binarity indicators, and 4 of these 26 stars host hot WD companions identified via SED modeling, corresponding to an overall observed binary fraction of $\sim$38\%. For comparison, in Trumpler~5, only 1 out of 10 BSSs (10\%) hosts a hot companion \citep{Chand2026}, while other OCs in the UVIT Open Cluster Survey (UOCS) report fractions of 20--60\% for BSSs with UV excess indicative of mass-transfer formation \citep{2022MNRAS.511.2274V, 2024MNRAS.52710335P}. The relatively low detection rate in our sample and in Trumpler~5 can be attributed to observational limitations, including sensitivity primarily to hot companions ($T_\mathrm{eff} > 11,000$~K), non-uniform differential reddening affecting UV flux, and the possibility of undetected Case-A or Case-B mass-transfer systems with cooler companions. Consequently, the observed binary fraction of $\sim$38\% should be considered a lower limit, given the limited spectroscopic coverage and sensitivity, which are primarily to hot companions, and serves as a useful baseline for comparison with other OCs.

\section{Structural Properties and Dynamical State}
\label{sec: rdp}
To derive the internal structural characteristics of Tombaugh~2, we analyzed its radial stellar density distribution using {\it Gaia} astrometric data. The radial surface density profile (RDP) was constructed using only high-probability cluster members (p $\ge$ 0.7). RDP, defined as the projected stellar surface density as a function of radial distance from the cluster center, was constructed to examine the structural properties of the cluster.

The RDP was constructed using an equal-area annular binning scheme,  in which each annulus subtends an equal projected area on the sky. This ensures a more uniform sampling of the stellar density field and yields comparable Poisson uncertainties across bins.  The number of bins was chosen such that each annulus contains at least $N_i \geq 20$ member stars, balancing statistical reliability with radial resolution. In the innermost regions where the stellar density is highest, adjacent bins were merged as necessary to satisfy this criterion. The uncertainties shown in the RDP correspond to Poisson errors propagated to surface density units ($\sigma = \sqrt{N}/A$), where $A$ is the area of each annulus. Despite maintaining a minimum number of stars per bin, uncertainties in the innermost regions are relatively larger due to area normalization and intrinsic spatial fluctuations in stellar density.

The stellar surface density,  $\Sigma(r_i)$, at a projected distance $r_i$ from the cluster center was obtained by counting the number of stars ($N_i$) enclosed within concentric rings of increasing radius and dividing by the corresponding annular area ($A_i$), such that  $\Sigma(r_i) = N_i / A_i$. The observed radial profile was fitted with the empirical surface-density model introduced by \citet{King62}, which effectively describes the variation in stellar density from the cluster core to its outer tidal boundary. The King function characterizes the surface density as:
\begin{equation}
{ \Sigma(r) = \Sigma_0 \left[ \frac{1}{\sqrt{1 + (r/r_c)^2}} - \frac{1}{\sqrt{1 + (r_t/r_c)^2}} \right]^2 + \Sigma_{bg}},
\label{eq:king_model}
\end{equation}
where  $\Sigma_0$ denotes the surface density normalization factor, $r_c$ the core radius (the distance where $\Sigma = 0.5\,\Sigma_0$), $r_t$ the tidal radius beyond which the stellar density merges with the Galactic background, and  $\Sigma_{bg}$ the residual field density along the line of sight.

Since the profile is constructed using only high-probability cluster members, the background term is expected to be negligible ($\Sigma_{bg} \approx 0$). Nevertheless, we keep $\Sigma_{bg}$ as a free parameter to account for residual contamination or membership incompleteness, and find it to be consistent with zero within uncertainties.

Parameter optimization was performed  by minimizing a chi-square statistic defined as:
\begin{equation}
\ln L = -\frac{1}{2} \sum_i \left(\frac{\Sigma_i - \Sigma_{i,\rm model}}{\sigma_{\Sigma_i}}\right)^2
\label{eq:log_likelihood}
\end{equation}
where $\Sigma_i$ is the observed surface density within each annulus, $\sigma_{\Sigma_i}$ the Poisson uncertainty, and $\Sigma_{i,{\rm model}}$ the corresponding King model prediction.

For the numerical implementation, the \texttt{emcee} package \citep{emcee} was used, with 100 random walkers and  5000 iterations. The initial 500 steps of each chain were discarded as burn-in to ensure that the sampling reflected the stationary distribution. Uniform priors were adopted for all parameters. Specifically, we adopted uniform priors over physically motivated ranges: $\Sigma_0 \in [0, \Sigma_{\rm max}]$, $r_c \in [0, r_{\rm max}]$, $r_t \in [r_c, r_{\rm max}]$, and $\Sigma_{bg} \in [0, \epsilon]$, where $\epsilon$ is a small value reflecting the expected negligible background level due to the use of high-probability members only.

\begin{figure}[ht]
\centering
\includegraphics[width=1\linewidth]{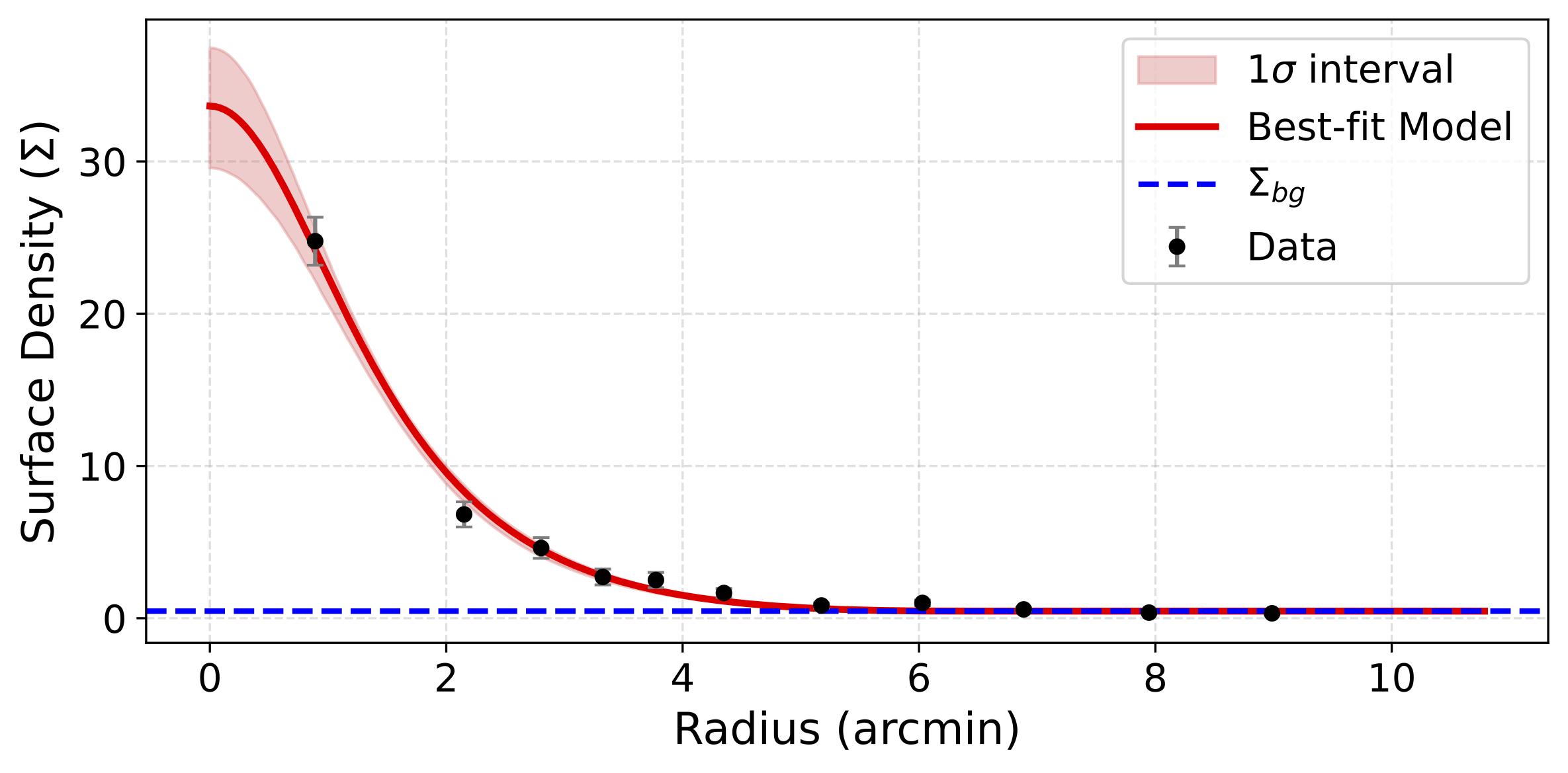}\\
\includegraphics[width=1\linewidth]{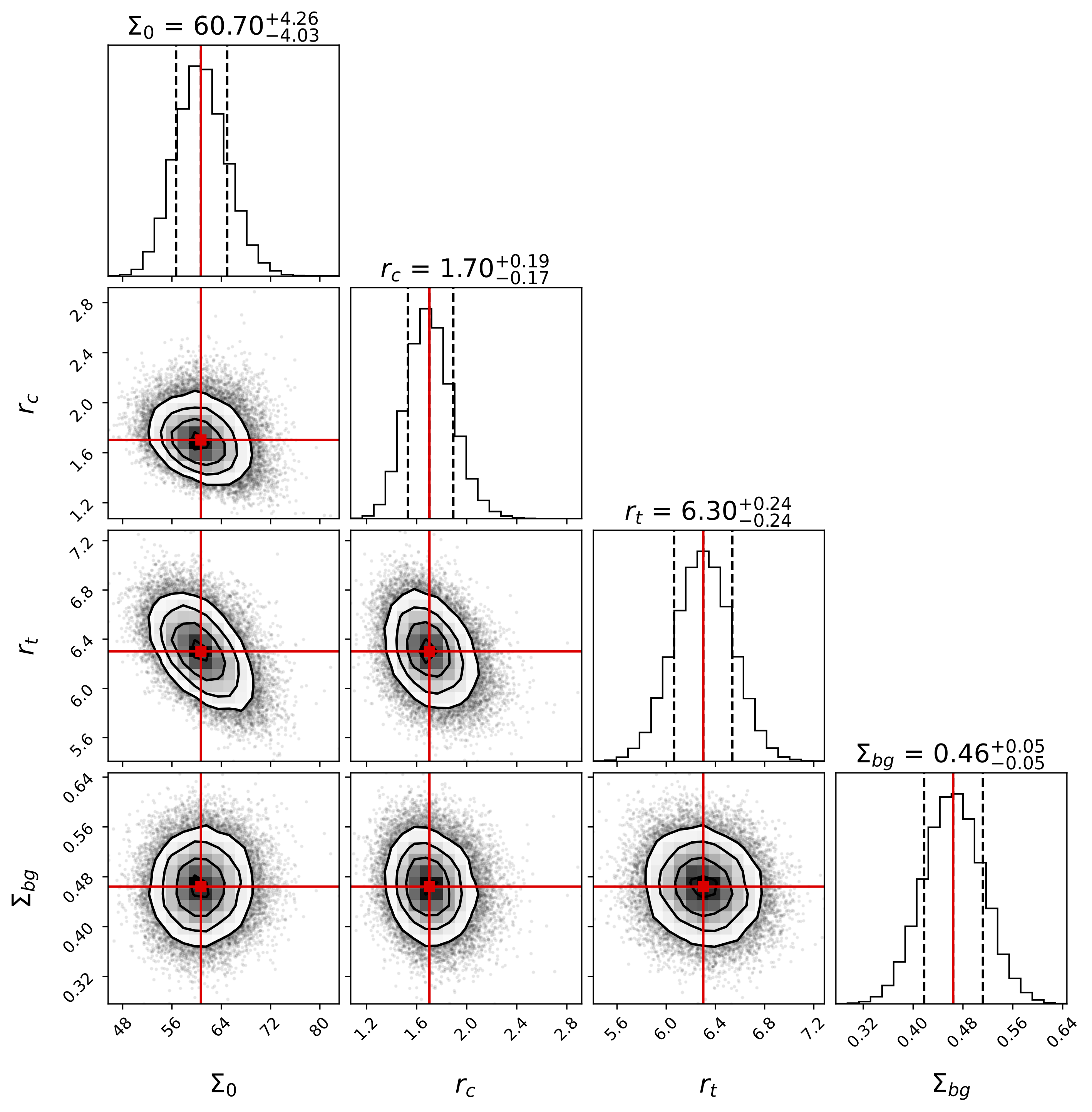}
\caption{Radial stellar density profiles. The \textbf{upper} panel shows the observed stellar surface densities (black points with Poisson uncertainties) as a function of radial distance from the cluster center, along with the best-fitting \citet{King62} models (solid red curve). The blue dashed line indicates the estimated background field density. The shaded regions represent the $1\sigma$ confidence intervals of the model fits derived from the posterior distributions. The \textbf{lower} panel shows the corresponding corner plot illustrating the posterior probability distributions and parameter correlations for the King-model parameters.}
\label{fig:RDP_fits}
\end{figure}

The convergence of the Markov chains was verified using the \citet{gelman1992} $\hat{R}$ diagnostic, which consistently yielded $\hat{R} \leq 1.2$, confirming stable convergence. The resulting fit and posterior distributions are illustrated in Figure~\ref{fig:RDP_fits}, and the corresponding parameters are listed in Table~\ref{tab:King_results}. Our estimated parameters, such as the core radius, are consistent with \citet{rao2023determination}; however, the tidal radius is smaller in our study, reflecting stricter membership selection and the treatment of the background field.

\begin{table}[!ht]
\renewcommand{\arraystretch}{1.2}
\caption{King-profile structural parameters of Tombaugh~2 derived from its radial density distribution. Parameters include surface density normalization ($\Sigma_0$), background density ($\Sigma_{bg}$), core radius ($r_c$), and tidal radius ($r_t$), with $1\sigma$ uncertainties.}
\label{tab:King_results}
\centering
\setlength{\tabcolsep}{4pt}
\begin{tabular}{cccc}\hline\hline
$\Sigma_0$ & $\Sigma_{bg}$ & $r_c$ & $r_t$  \\
  \multicolumn{2}{c}{(stars arcmin$^{-2}$)} & \multicolumn{2}{c}{(arcmin)} \\
\hline
 {$60.70^{+4.26}_{-4.03}$ }
& {$0.46^{+0.05}_{-0.05}$ }
& {$1.20^{+0.19}_{-0.17}$ }
& {$6.30^{+0.24}_{-0.24}$ }\\
\hline
\end{tabular}
\end{table}

Although the cluster's core radius is relatively small, it remains less compact than typical evolved OCs, even when considering the compactness parameter. The cluster's RDP indicates a low-density system with modest concentration, characteristic of intermediate-age, dynamically evolved OCs that have experienced significant two-body relaxation and partial mass segregation \citep{Bonatto2005, Piskunov2007}. Nevertheless, the stellar density is still too low for collisions to play a major role, implying that the BSS population likely arises from post-interaction binaries rather than collisional formation channels.

Therefore, Tombaugh~2 is better described as a dynamically evolved open cluster rather than a globular cluster-like collisional system. In such a low-density, dynamically relaxed environment, long-period interacting binaries are expected to survive on Gyr timescales. This dynamical expectation is directly testable through RV monitoring, which we exploit in Section~\ref{sec:RV_analysis} to assess the binary nature of the BSS population.

\section{Dynamical Environment and Implications for BSS Formation}
\label{sec: bss_dynamics}

The presence of WD companions among the BSSs in Tombaugh~2 provides strong and direct evidence that mass transfer is the dominant formation channel in this cluster. Similar conclusions have been reached for other old OCs, such as NGC~7789 \citep{2024BSRSL..93..250V} and NGC~188 \citep{Subramaniam2016}, where UV excesses in the SEDs of BSSs clearly trace past binary interaction. These results underline that the formation of BSSs in dynamically evolved OCs is closely linked to the long-term survival and evolution of interacting binaries.

The results from our RV analysis (Section \ref{sec:RV_analysis}) further reinforce this picture by providing independent kinematic confirmation of the binary nature of these systems. While the limited time baseline and the small number of epochs in our spectroscopic dataset do not permit the derivation of full Keplerian orbital solutions or precise mass ratios for individual systems, the statistically significant velocity scatter ($\sigma_{obs} \gg \langle \epsilon \rangle$) observed in the BSS sample is an unambiguous signature of binarity. This high fraction of RV variables is complementary to the detection of hot WD companions via SED modeling. Together, the photometric and spectroscopic evidence suggests that the BSS population in Tombaugh~2 is largely composed of post-interaction systems formed through binary mass transfer.

To place these findings into a broader Galactic context, we examine the kinematic and orbital properties of Tombaugh~2. From a theoretical perspective, a cluster's Galactic orbit governs the strength and frequency of external tidal perturbations it experiences over its lifetime. Clusters on eccentric or dynamically heated orbits, particularly those venturing into the inner disk, are exposed to strong tidal forces, disk shocking, and encounters with dense molecular clouds. Such conditions can lead to significant mass loss and efficient disruption of binary systems, as observed in clusters like NGC~752 \citep{2024A&A...686A.215L}. In contrast, clusters evolving in the outer disk on nearly circular orbits are subject to a much quieter tidal environment, which favours the long-term survival of binaries undergoing Case~A or Case~B mass transfer, as suggested in the analysis performed by \citet{2024ApJ...961..251Y}, for the open cluster NGC 2420.

Although BSSs are formed through internal binary evolution, the Galactic orbit of an open cluster can play an important supporting role by shaping the external tidal environment in which these binaries evolve. Clusters following eccentric or inner-disk orbits are repeatedly exposed to disk crossings, tidal shocks, and encounters with giant molecular clouds, all of which can enhance mass loss and progressively erode the cluster's binary population over time \citep{BaumgardtMakino2003, Lamers2005}. Indeed, independent observational and theoretical studies evaluating the overall binary fractions of OCs, distinct from BSS-specific analyses, confirm this dynamical picture; clusters subjected to stronger tidal perturbations and higher encounter rates in the inner Galactic disk exhibit systematically lower binary survival fractions than those evolving in the less disturbed outer disk \citep{2010MNRAS.401..577S, 2011MNRAS.417.1684M, 2023A&A...676A..56D}. By contrast, clusters on nearly circular orbits in the outer disk evolve in a much calmer tidal environment, allowing interacting binaries to survive for several Gyrs and complete the evolutionary pathways leading to blue straggler formation. Observational studies of old, dynamically evolved OCs support this picture, indicating that such quiescent environments tend to host BSS populations that are likely dominated by mass-transfer and merger products rather than by stellar collisions \citep{Gosnell2015, Subramaniam2016}. In this broader context, the Galactic orbit of Tombaugh~2 provides a natural and physically consistent backdrop for the properties of its BSS population.

Using the present-day astrometric and radial-velocity data, we computed the space velocity of Tombaugh~2 relative to the Local Standard of Rest (LSR), adopting the Solar motion from \citet{Coskunoglu2011}. The resulting velocity components and total space velocity are listed in Table~\ref{Tab:dyn_params_new}. Tombaugh~2 exhibits a modest lag with respect to Galactic rotation and a relatively large radial motion, yielding a total space velocity of $\sim 104$~km~s$^{-1}$. This kinematic behaviour is consistent with an old disk population \citep{Bensby2003, Robin2003} and indicates that the cluster has experienced a dynamically evolved, but not necessarily strongly disruptive, Galactic history.

\begin{table}[ht]
\centering
\renewcommand{\arraystretch}{1.1}
\setlength{\tabcolsep}{8pt}
\caption{Derived kinematic and orbital parameters for Tombaugh~2. The listed quantities include the Galactocentric distance ($R_{\mathrm{gc}}$), the space velocity components corrected for the LSR ($U$, $V$, $W$), the total space velocity ($S_{\mathrm{LSR}}$), the maximum distance from the Galactic plane ($Z_{\mathrm{max}}$), apogalactic distance ($R_{\mathrm{a}}$), perigalactic distance ($R_{\mathrm{p}}$), mean orbital radius ($R_{\mathrm{m}}$), orbital eccentricity ($e$), and orbital period ($T_{\mathrm{p}}$).}
\begin{tabular}{l c}
\hline\hline
\multicolumn{2}{c}{\textbf{Kinematic Parameters}} \\
\hline
$R_{\rm gc}$ (kpc)            & 13.48~$\pm$~1.09 \\
$U_{\rm LSR}$ (km~s$^{-1}$)  & $-95.61$~$\pm$~16.75 \\
$V_{\rm LSR}$ (km~s$^{-1}$)  & $-40.72$~$\pm$~4.25 \\
$W_{\rm LSR}$ (km~s$^{-1}$)  & $-0.50$~$\pm$~4.55 \\
$S_{\rm LSR}$ (km~s$^{-1}$)  & 103.92~$\pm$~17.87 \\
\hline
\multicolumn{2}{c}{\textbf{Orbital Parameters}} \\
\hline
$Z_{\rm max}$ (kpc)           & 0.83~$\pm$~0.21 \\
$R_{\rm p}$ (kpc)             & 12.40~$\pm$~1.33 \\
$R_{\rm a}$ (kpc)             & 13.77~$\pm$~1.67 \\
$R_{\rm m}$ (kpc)             & 13.09~$\pm$~1.50 \\
$e$                           & 0.05~$\pm$~0.01 \\
$T_{\rm p}$ (Myr)             & 385~$\pm$~49 \\
\hline
\end{tabular}
\label{Tab:dyn_params_new}
\end{table}

\begin{figure}[ht]
\centering
\includegraphics[width=0.95\linewidth]{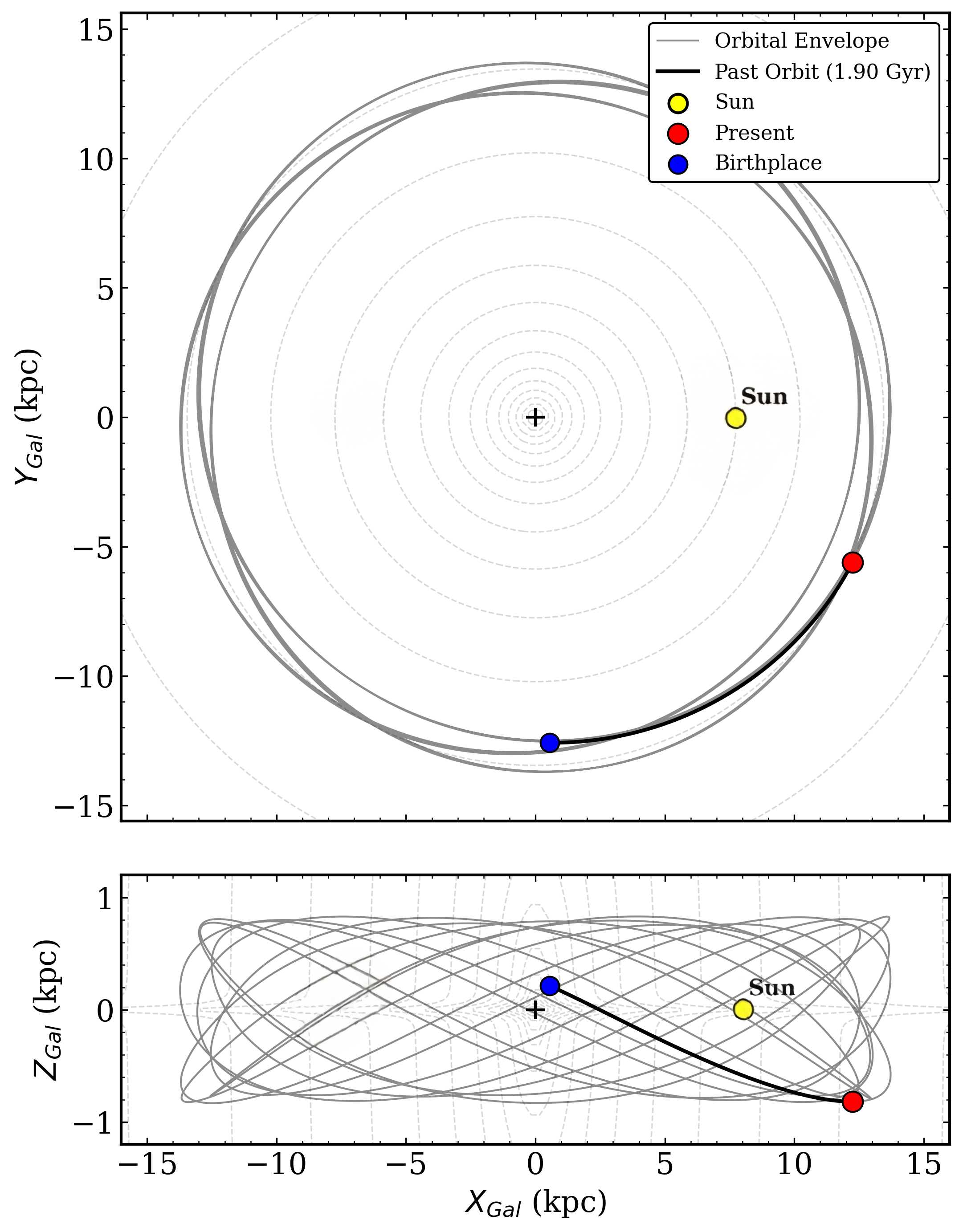}
\caption{Projected Galactic orbit of the cluster in the $X_{\rm Gal}$–$Y_{\rm Gal}$ (top) and $X_{\rm Gal}$–$Z_{\rm Gal}$ (bottom) planes. The solid black curve shows the past orbit integrated over the last 1.74~Gyr, while the faint structure indicates the orbital envelope. Red and blue circles mark the present-day position and the inferred birth location of the cluster, respectively, and the yellow symbol marks the Sun. Owing to the low orbital eccentricity and the assumed axisymmetric Galactic potential, the orbit appears spatially confined and quasi-periodic over the adopted time interval. The grey dashed contours show the logarithmic mass density of the Milky Way in the \texttt{MWPotential2014} model.}
\label{fig:orbit}
\end{figure}

To further characterize the dynamical environment, we derived the Galactic orbit of Tombaugh~2 using its complete six-dimensional phase-space information. Orbit integrations were performed with the \texttt{galpy} package \citep{Bovy2015} in the axisymmetric \texttt{MWPotential2014}. The resulting orbital parameters are summarized in Table~\ref{Tab:dyn_params_new}, and the orbit projections are shown in Figure~\ref{fig:orbit}.

The integrated orbit shows that Tombaugh~2 resides in the outer Galactic disk at a Galactocentric distance of $R_{\rm GC} \approx 13$~kpc and follows a low-eccentricity orbit ($e < 0.1$). Such an orbit implies that the cluster has largely avoided strong tidal shocks and repeated passages through the dense inner disk. Its vertical excursion of roughly 1~kpc above the Galactic plane further reduces the frequency of encounters with massive molecular clouds, supporting the view that Tombaugh~2 has evolved in a relatively tranquil dynamical environment.

We note that the apparent confinement of the orbit to a relatively small spatial region over $\sim$1.7~Gyr is a natural consequence of the cluster's low orbital eccentricity and the use of an axisymmetric Galactic potential. In such a potential, nearly circular disk orbits are expected to be quasi-periodic and to retrace similar paths over long integration times. The integration interval was therefore chosen to adequately visualize the orbital envelope rather than to imply significant radial migration.

This stable outer-disk orbit provides a natural dynamical framework for the SED-based results. The weak external tidal field implied by the orbit strongly favours the survival of interacting binaries over Gyr timescales, allowing them to complete their internal evolutionary pathways and produce BSS+WD systems. In this context, collisional BSS formation is expected to be inefficient, while mass transfer naturally emerges as the dominant channel. A similar interpretation has been proposed for NGC~2420 by \citet{2024ApJ...961..251Y}, which occupies a comparable region of the Galaxy.

In contrast to clusters such as NGC~752, where strong tidal perturbations have been linked to significant binary disruption \citep{2024A&A...686A.215L}, the orbital properties of Tombaugh~2 indicate that its BSS population is shaped primarily by internal cluster dynamics. The observed BSSs are therefore best understood as the outcome of long-lived binary evolution, rather than as products of external Galactic forcing.

\section{Conclusion $\&$ Summary}
\label{sec: conclusion_summary}

We have carried out a multiwavelength investigation of the open cluster Tombaugh~2, combining UV, optical, and infrared photometry with Gaia DR3 astrometry and multi-epoch spectroscopy. This integrated analysis enables a self-consistent characterization of the cluster's structural and orbital properties and, in particular, the evolutionary nature of its blue straggler population. Our results show that while the cumulative radial distribution indicates only a mild central concentration of BSSs within the inner regions, their distribution at larger radii becomes comparable to that of RGB stars. This reveals that the cluster is dynamically intermediate and not yet fully relaxed. Furthermore, the detection of UV excesses and hot compact companions provides direct evidence that binary mass transfer plays a significant role in the formation of blue stragglers in this low-density environment. The main findings are summarized as follows:

\begin{enumerate}
\item  Using a GMM applied to \textit{Gaia}~DR3 astrometric data, we identified 562 probable members in Tombaugh~2. The derived mean proper motions are consistent with values reported by \citet{Cantat-Gaudin20, hunt2023improving}, supporting the reliability of the adopted membership selection.

\item Isochrone fitting in the Gaia CMD yields an age of $1.74 \pm 0.1$~Gyr for Tombaugh~2, indicating that the cluster is of intermediate age in the outer Galactic disk. The corresponding distance is $7.1 \pm 0.5$~kpc, consistent with previous photometric and astrometric studies. The cumulative radial distribution indicates only mild central concentration of BSSs, with no strong evidence of full mass segregation. This suggests that the cluster is in a dynamically intermediate state, consistent with \citet{rao2023determination}.

\item Structural parameters obtained from King-profile fitting reveal a core radius of $r_c = 1.2'$ and a tidal radius of $r_t = 6.3'$ for Tombaugh~2. The derived concentration parameter ($c \approx 0.72$) indicates a moderate level of central concentration, comparable to dynamically young or intermediate-age stellar systems, rather than a highly concentrated or dynamically evolved cluster.

\item The SEDs of confirmed cluster members were modeled across the UV–IR range using VOSA. Several BSSs display prominent UV excesses in the \textit{Swift}/UVOT bands that are inconsistent with single-star atmospheric models. Two-component SED fitting, incorporating hot WD companions, successfully reproduces the observed fluxes, indicating that a significant fraction ($\sim$30\%) of the BSS population likely hosts post-mass-transfer binaries. This fraction is comparable to values reported in other UV-based studies of OCs (typically $\sim$30–50\%), while optical or spectroscopic studies often report lower fractions due to their reduced sensitivity to hot companions. Since UV observations preferentially detect recently formed systems, this fraction should be regarded as a lower limit.

\item SED modeling reveals a heterogeneous population of BSS companions. The nine binaries host hot secondaries spanning $T_{\rm eff}\approx 15$–70~kK with radii $R\approx 0.04$–$0.28\,R_\odot$. These include inflated pre-ELM / pre-He WD candidates with radii $0.13$–$0.28\,R_\odot$ as well as young, compact hot WDs ($T_{\rm eff}\geq 30$~kK, $R\sim 0.04$–$0.08\,R_\odot$). The BSS primaries exhibit $T_{\rm eff}\simeq 7{,}500$–$9{,}000$~K and $R\simeq 1.5$–$2.3\,R_\odot$, consistent with rejuvenated mass-transfer products. The full sample consists of 19 single-component and 9 BSS+WD systems.

\item The combination of UV excess detections, SED-derived stripped companions, and multi-epoch RV variability provides strong, multi-faceted evidence that binary mass transfer is a dominant formation pathway for BSSs in this low-density environment. The observed RV variations, in several cases exceeding $100$~km~s$^{-1}$ over timescales of weeks, are significantly larger than the internal measurement uncertainties ($\sigma_{\rm obs} \gg \langle \epsilon \rangle$), confirming a high binary fraction among the BSS population. These independent photometric and spectroscopic diagnostics consistently support the presence of interacting or post-interaction binary systems, although contributions from merger channels and dynamically modified binaries cannot be ruled out.

\item Orbital integrations using \texttt{galpy} within the \texttt{MWPotential2014} Galactic model reveal that Tombaugh~2 follows a nearly circular, thin-disk orbit with low eccentricity ($e \approx 0.05$–$0.08$) and modest vertical excursions ($Z_{\max} < 1$~kpc). The derived orbital period ($T_p \sim 400$~Myr) and mean Galactocentric radius ($R_{\rm m} \approx 13$–14~kpc) confirm that the cluster is a long-lived member of the outer thin disk, unaffected by strong perturbations or halo interactions.

\item The combination of UV diagnostics, SED modeling, and orbital dynamics presents a coherent picture in which Tombaugh~2 is a representative example of an intermediate-age, dynamically intermediate open cluster in the outer Milky Way. The evidence for multiple BSS~+~WD systems provides a direct link between binary evolution and the cluster's dynamical state, reinforcing the idea that binary mass transfer is the primary channel for BSS formation in low-density Galactic environments.
\end{enumerate}

In summary, this study establishes Tombaugh~2 as a benchmark for understanding the interplay between stellar dynamics, binary evolution, and cluster survival in the outer Galactic disk. The presence of UV-bright blue straggler binaries underscores the critical role of binary-interaction processes in shaping the stellar populations of intermediate OCs. It provides valuable constraints on population synthesis models of the Galactic disk. Our results suggest that UV observations are essential for uncovering the true binary nature of BSS populations in intermediate- old OCs, as analyses restricted to optical and infrared wavelengths alone can significantly underestimate the fraction of post-mass-transfer systems. Future time-domain UV observations and spectroscopy will be essential for confirming the binary nature of these systems and further constraining their evolutionary histories.

\begin{acknowledgments}
 We sincerely thank the anonymous referee for a thorough review and constructive suggestions that greatly improved the clarity and quality of the manuscript. Ing-Guey Jiang acknowledges support from the National Science and Technology Council (NSTC), Taiwan, under grants NSTC 113-2112-M-007-030 and NSTC 114-2112-M-007-029. This work uses data from the European Space Agency's Gaia mission, processed by the Gaia Data Processing and Analysis Consortium (DPAC). Ultraviolet data were obtained from the Neil Gehrels Swift Observatory using calibrated Level-3 UVOT point-source catalogs (\citep{Siegel2019}). Optical photometry was taken from Pan-STARRS1 and the SkyMapper Southern Survey. Near- and mid-infrared data were obtained from 2MASS and WISE. Spectroscopic data were retrieved from the ESO Science Archive Facility based on observations obtained with VLT/FLAMES under Programme ID 076.D-0220. This publication makes use of VOSA, developed under the Spanish Virtual Observatory (https://svo.cab.inta-csic.es) project funded by MCIN/AEI/10.13039/501100011033/ through grant PID2020-112949GB-I00. VOSA has been partially updated by using funding from the European Union's Horizon 2020 Research and Innovation Programme, under Grant Agreement nº 776403 (EXOPLANETS-A)
\end{acknowledgments}



\bibliography{References}{}
\bibliographystyle{aasjournalv7}



\end{document}